\documentclass[a4paper,11pt]{article}
\usepackage{pos}
\usepackage{graphicx}
\usepackage{epsfig}
\usepackage{bm}
\usepackage{mathrsfs}  
\usepackage{enumitem}

\newcommand{\postscript}[2]{\setlength{\epsfxsize}{#2\hsize}
   \centerline{\epsfbox{#1}}}

\title{Landscape, Swampland, and Extra Dimensions}

\author[a,b,c]{Luis A. Anchordoqui}
\author[d,e]{Ignatios Antoniadis}
\author[f,g]{Dieter L\"ust}

\affiliation[a]{Department of Physics and Astronomy,  Lehman College, City University of
  New York, NY 10468, USA
}

\affiliation[b]{Department of Physics,
 Graduate Center, City University
  of New York,  NY 10016, USA
}

\affiliation[c]{Department of Astrophysics,
 American Museum of Natural History, NY
 10024, USA
}

\affiliation[d]{High Energy Physics Research Unit, Faculty of Science, Chulalongkorn University, Bangkok 1030, Thailand}

\affiliation[e]{Laboratoire de Physique Th\'eorique et Hautes \'Energies - LPTHE
Sorbonne Universit\'e, CNRS, 4 Place Jussieu, 75005 Paris, France
}

\affiliation[f]{Max--Planck--Institut f\"ur Physik,  
 Werner--Heisenberg--Institut,
80805 M\"unchen, Germany
}

\affiliation[g]{Arnold Sommerfeld Center for Theoretical Physics, 
Ludwig-Maximilians-Universit\"at M\"unchen,
80333 M\"unchen, Germany
}

\emailAdd{luis.anchordoqui@lehman.cuny.edu}
\emailAdd{antoniad@lpthe.jussieu.fr}
\emailAdd{luest@mpp.mpg.de}

\abstract{By combining swampland conjectures with observational data, it was recently suggested that the cosmological hierarchy problem (i.e. the smallness of the dark energy in Planck units) could be understood as an asymptotic limit in field space, corresponding to a decompactification of one extra (dark) dimension of a size in the micron range. In these Proceedings we examine the fundamental setting of this framework and discuss general aspects of the effective low energy theory inherited from properties of the overarching string theory. We then explore some novel phenomenology encompassing the dark dimension by looking at potential dark matter candidates, decoding neutrino masses, and digging into new cosmological phenomena.}

\FullConference{Based on talks given by the authors on several
  conferences in 2023 and 2024.
}

\begin{document}
\maketitle

\section{Introduction}\label{s:intro}

The challenge for a fundamental theory of nature is to describe both particle physics and cosmology. Accelerator experiments and
cosmological observations provide complementary information to constrain the
same theory. We have long known that only about 4\% of the content of
the universe is ordinary baryonic matter; the remainder is dark matter
($\sim 22\%$) and dark energy ($\sim 74\%$). The $\Lambda$CDM model,
in which the expansion of the universe today is dominated by the
cosmological constant $\Lambda$ and cold dark matter (CDM), is the
simplest model that provides a reasonably good account of all
astronomical and cosmological observations~\cite{ParticleDataGroup:2022pth}.

The cosmological evolution is described by Einstein's field equation,
\begin{equation}
 {\cal  R}_{\mu \nu} - \frac{1}{2} \ g_{\mu \nu} \ {\cal R}
 + g_{\mu \nu} \ \Lambda = \frac{8 \pi G}{c^4} \ T_{\mu \nu}   \,,
\label{Efe}
\end{equation}
where ${\cal R}_{\mu \nu}$ and ${\cal R}$ are respectively the Ricci tensor and scalar,
$g_{\mu \nu}$ is the metric tensor, $T_{\mu \nu}$ is the energy
momentum tensor, Greek subscripts run from 0 to 3, and $G$ is Newton's gravitational constant. The cosmological constant encapsulates 
two length scales: the size
of the observable Universe $[\Lambda] = L^{-2}$ and of the dark
energy $[\Lambda/G \times
c^3/\hbar] = L^{-4}$. The observed value of the cosmological constant  $\Lambda_{\rm obs} \simeq 0.74 \times 3
H_0^2/c^2 \simeq 1.4 \times (10^{26}~{\rm m})^{-2}$ gives a characteristic
length of dark energy $\simeq 85~\mu{\rm m}$, where we
have adopted the recent measurement of the Hubble constant $H_0 \simeq
73~{\rm km/s/Mpc}$ by the SH0ES team~\cite{Riess:2021jrx,Murakami:2023xuy}.

The $SU(3)_C \otimes SU(2)_L
\otimes U(1)_Y$ Standard Model (SM) of strong and electroweak
interactions encapsulates our current best understanding of physics at
the smallest distances ($< 10^{-21}~{\rm
  m}$) and highest energies (center-of-mass energies $\sqrt{s}
\sim 14~{\rm TeV}$). Even though the SM  continues to survive all
experimental tests at accelerators~\cite{ParticleDataGroup:2022pth},
it is widely considered to be an effective theory. Besides missing gravity, the SM does not include a mechanism for
giving neutrinos their masses, and does not incorporate dark matter or
dark energy. Moreover, if
we compare the strength of gravity to that of the SM interactions we find that
\begin{equation}
  G M_H^2/(\hslash c) = (M_H/M_p)^2 \approx 10^{-34} \ll 1 \,,
\end{equation}
where $M_H$ is the Higgs mass and $M_p$ is the Planck mass. 

Leaving aside for the moment the SM downsides, a question to ask ourselves is:
{\it why is the gravitational interaction between SM particles so much
  weaker than the other SM interactions?} On the flip side, {\it why is the Planck mass so huge relative to the SM
  or dark energy scales?}

A way
to explain hierarchies in fundamental physics is via the size of extra dimensions which are necessary ingredients for consistency of string 
theory. 
Indeed, if the size of the extra dimensions is large compared to the fundamental (string) length, the strength of gravitational interactions becomes strong at distances larger than the actual four-dimensional (4D) Planck 
length~\cite{Arkani-Hamed:1998jmv, Antoniadis:1998ig}. As a result, the string scale is detached from the Planck mass consistently with all experimental bounds if the observable universe is localized in the large compact space~\cite{Antoniadis:1998ig}.
By combining swampland conjectures with observational data, it was recently suggested~\cite{Montero:2022prj} that the cosmological hierarchy problem (i.e. the smallness of the dark energy in Planck units) could be understood as an asymptotic limit in field space, corresponding to a decompactification of one extra (dark) dimension of a size in the micron range. 
In addition in cosmology, such an extra fifth dimension also provides a nice explanation for the 60 e-foldings in the course of cosmic inflation~\cite{Anchordoqui:2022svl,Anchordoqui:2023etp}.

In these Proceedings we summarize the state-of-the-art in this subject
area, and discuss future research directions. We begin by reviewing the emergence of a new paradigm of quantum gravity.

\section{The Landscape and the Swampland}

\begin{figure}[htb!]
   \begin{minipage}[t]{0.48\textwidth}
    \postscript{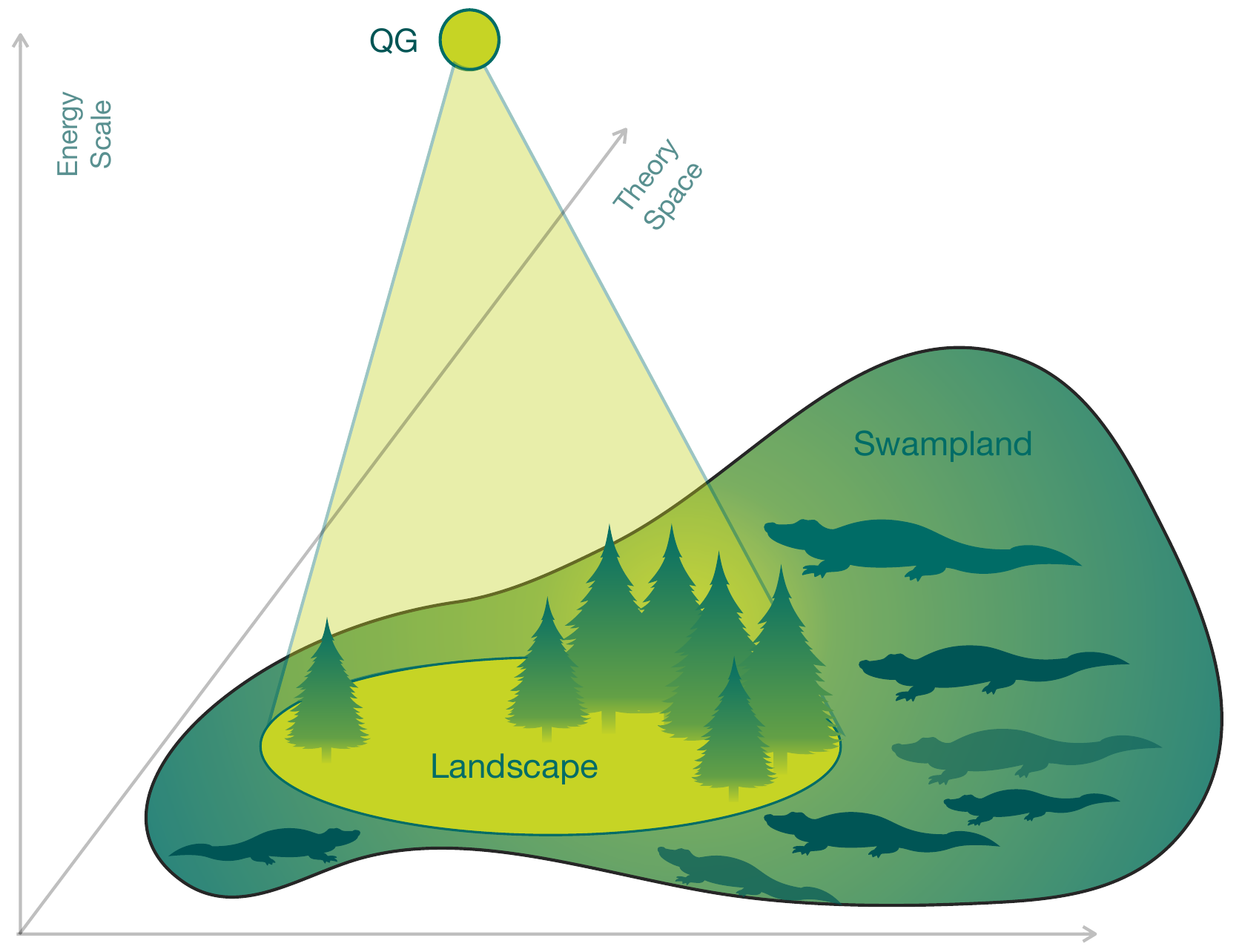}{0.95}
  \end{minipage}
\begin{minipage}[t]{0.48\textwidth}
    \postscript{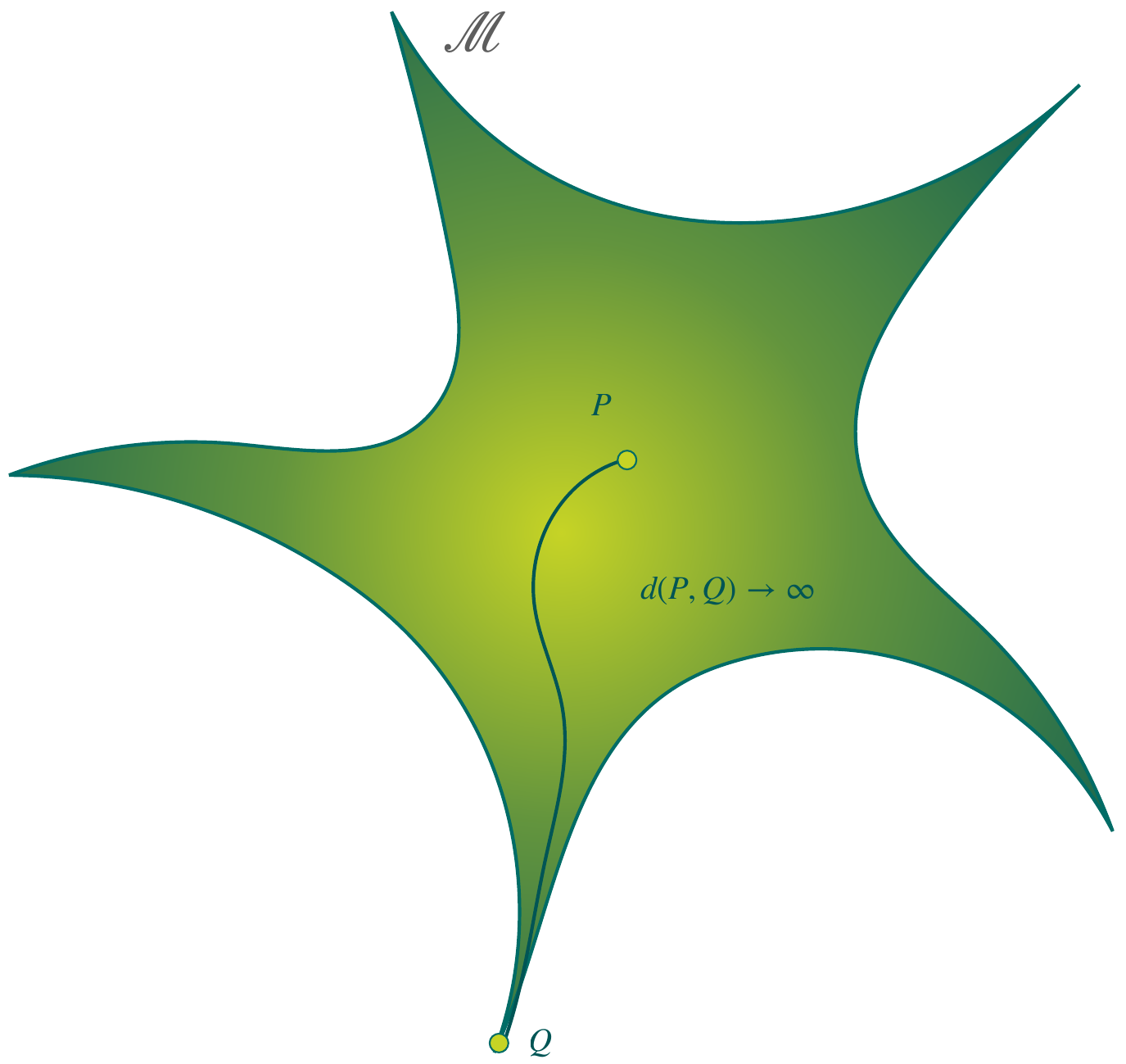}{0.8}
  \end{minipage}
  \caption{{\it Left.} The space of EFTs that are a low energy limit
    of quantum gravity forms a cone in theory space,
    because the swampland constraints become stronger at high
    energies. {\it Right.} Moduli space $\mathscr{M}$ with an infinite distance limit
    between points $P$ and $Q$. From Ref.~\cite{Makridou:2023wkb}. \label{fig:1}}
\end{figure}

The Swampland Program seeks to amplify our understanding of the
fundamental constraints accompanying a consistent theory of quantum
gravity (QG)~\cite{Vafa:2005ui}. The basic thought is that an effective
field theory (EFT) might seem consistent as a stand-alone theory in
the IR (anomaly free), but coupling the theory to gravity in the UV
may uproot its consistency. Actually, given the scarcity of consistent
EFTs coupled to QG, it is easier to rule inconsistent theories out. 
The goal is therefore to circumscribe the
subset of 4D EFTs that can be UV-completed to a QG theory and are
said to belong in the landscape from the complementary subset of
theories that do not admit such a completion and are relegated to the
swampland. This is done by enumerating criteria classifying the
properties that an EFT must
satisfy in order to enable a consistent completion into QG.
As the energy increases and we get closer to the QG scale
the swampland criteria provide stronger constraints on the boundary
that separates the swampland from the landscape. This implies that the
space of UV-consistent EFTs encircles a conical-shape structure that
is cartooned in the left panel of Fig.~\ref{fig:1}.

The swampland criteria have evolved to some set of conjectures, which
can be used as new guiding principles to construct compelling
UV-completions of the SM. There are many swampland conjectures in the literature; 
actually, too many to be listed here and so readers are
referred to comprehensive reviews~\cite{Palti:2019pca,vanBeest:2021lhn,Grana:2021zvf,Agmon:2022thq}. Conjectures linked to towers of
states, however, deserve some attention because they are closely related to the topics
we will explore in this proceedings.

Consider a gravitational EFT  with a moduli space $\mathscr{M}$ (i.e. a
space parameterized by the massless scalar fields in the theory) and
whose metric is given by the kinetic terms of the scalar
fields.\footnote{In the presence of a mechanism of moduli stabilization, the moduli space is discretized 
corresponding to a finite landscape of vacua which is assumed to be large enough for the following arguments to hold.}
The {\bf distance conjecture} (DC) states~\cite{Ooguri:2006in} that infinite distance limits $\Delta \phi \to \infty$, with
reference to geodesic field distance $\Delta \phi \equiv d(P,Q)$
illustrated in Fig.~\ref{fig:1}, are accompanied by an infinite tower of exponentially light states 
\begin{equation}
m (Q) \sim m (P) \ e^{-\lambda \Delta \phi} \, ,
\end{equation}
where distance and masses are measured in Planck units, and 
$\lambda$ is an order one positive constant.

The archetypal example to gain intuition about the DC is a theory
compactified on a circle. Taking $R$ to be the modulus controlling the
radius (or radion), it is straightforward to see
that the Kaluza-Klein (KK) modes have masses that scale as
$m_{n}^2= n^2/R^2$, with $n \in \mathbb{Z}$. Now, after
dimensional reduction of the gravitational piece of the action and the
corresponding  field redefinition to go into the Einstein-frame, the
kinetic term of the Lagrangian for the radion field takes the form
$\mathscr{L}_{{\rm kin}} \supset (\partial R)^2/ R^{2}$.
The distance in field space between points $R_{\rm i}$ and $R_{\rm f}$
is measured by the field space metric, given by $1/R^2$, and
is found to be
\begin{equation}
d(R_{\rm i}, R_{\rm f})=\int_{R_{\rm i}}^{R_{\rm f}}
\sqrt{\dfrac{1}{R^2}} \ dR= \left|\ln (R_{\rm f}/R_{\rm i}) \right|.
\end{equation}
As predicted by the DC, Taking the decompactification limit $R \to
\infty$ we see that the KK tower becomes light, $m_{n}\sim
e^{-d(R_{\rm i}, R_{\rm f} \rightarrow \infty)}$. In the opposite
limit at infinite distance $R \to 0$ the KK tower does not satisfy
the DC. However, in this limit there is another infinite tower that is
becoming light; namely, the winding modes (wrapping strings), which
present the same behaviour as the KK modes in the decompactification
limit. All in all, we arrive at a staggering conclusion: {\it to satisfy the DC in KK compactifications we need extended objects that can wrap some compact directions and become light in the limit in which they shrink to zero size}.

Before moving on, we bring up a refinement of the DC, which is known as the {\bf
 emergent string conjecture} (ESC) and states that any infinite distance limit is either a decompactification limit or a limit in which there is a weakly coupled string becoming tensionless~\cite{Lee:2019wij}.
 Recently it was argued \cite{Basile:2023blg,Bedroya:2024ubj}
that  bottom-up arguments  from black hole properties provide general evidence for the ESC.

The DC and the ESC can be generalized to other field space configurations beyond
the moduli space. For example, we can define a notion of distance
between different metric and flux configurations of anti-de Sitter spacetimes~\cite{Lust:2019zwm,Li:2023gtt,Basile:2023rvm,Palti:2024voy},
such that the flat space limit $\Lambda \to 0$ is located at
infinite distance in this metric configuration space. 
The related distance in the space of AdS vacua is then given as
\begin{equation}
d(\Lambda_{\rm i}, \Lambda_{\rm f})=\left|\ln (\Lambda_{\rm f}/\Lambda_{\rm i}) \right|,
\end{equation}
and  the {\bf anti-de Sitter distance conjecture}  (AdS-DC)
states~\cite{Lust:2019zwm} that any AdS vacuum has an infinite tower of states that
becomes light in the flat space limit $\Lambda \to 0$, satisfying
\begin{equation}
  m \sim |\Lambda|^\alpha \, ,
\end{equation}
with $\alpha$ a positive constant of order
one. Besides, under the hypothesis that
 this scaling behavior holds in de Sitter (dS) space, an unbounded
 number of massless modes also pop up in the limit $\Lambda \to
 0$~\cite{Lust:2019zwm}. A point worth noting at this juncture is 
that the AdS-DC modifies the EFT expectation
$\Lambda \sim \Lambda_0 + m^{1/\alpha}$, in which $\Lambda_0$ stands
for the contribution of the heavy modes. The vanishing of $\Lambda_0$
is argued to be intrinsically related to the modular invariance of
string theory (see also the discussion in~\cite{Anchordoqui:2023laz}).
Furthermore one can formulate a conformal field theory (CFT) distance conjecture about the spectrum of certain 
(holographically dual) conformal and world sheet field theories~\cite{Perlmutter:2020buo,Baume:2023msm,Ooguri:2024ofs,Aoufia:2024awo},
and one can also generalize it to topology change and non-geometric spaces at infinite distance~\cite{Demulder:2023vlo}.

In a  way, similar to the AdS-DC, a distance conjecture for black hole solutions in the EFT was put forward~\cite{Bonnefoy:2019nzv}.
Concretely the {\bf black hole entropy distance conjecture} (BHEDC) states that 
 the limit of 
infinite black entropy $S_{\rm BH}\rightarrow\infty$ 
is at infinite distance in 
 the space of BH metrics and
the associated BH entropy distance  can be expressed as 
\begin{equation}
d(S_{\rm BH, i}, S_{\rm BH, f})=\left|\ln (S_{\rm BH,  f}/S_{\rm BH, i}) \right|\, .
\end{equation}
Following the infinite distance conjecture,   there must be a corresponding mass scale with a tower of ``states", whose
 masses in Planck units are given as
\begin{equation}
m \sim({S_{\rm BH}})^{-\beta}\;,
\label{gendisc}
\end{equation}
where $\beta$ is a positive constant. 
Further discussion and more evidence for the BHEDC for extremal and non-extremal black holes was subsequently given in~\cite{Luben:2020wix,Cribiori:2022cho,Delgado:2022dkz,Angius:2023xtu}.
In addition, the BHEDC can be also applied for so called minimal black holes. Their minimal horizon size $L_{\rm min}$ serves as a 
working definition for the species scale $ \Lambda_{\rm sp}$~\cite{Dvali:2007hz,Dvali:2007wp,Dvali:2009ks,Dvali:2010vm,Dvali:2012uq}
and the corresponding species length $l_{\rm sp}$ in the EFT 
(see also the discussion in section 3.1 and section 8.):\footnote{Alternatively, $\Lambda_{\rm sp} = M_p/\sqrt{F_1}$ can be 
  identified with the scale at which $R^2$ corrections to the Einstein
  action become important, with $F_1 \simeq N$ being the one-loop topological
  string free energy~\cite{vandeHeisteeg:2022btw}. In \cite{Cribiori:2022nke} it was shown that this definition of the species agree with the one from minimal black holes for a specific type of charged black holes. 
  Further recent discussion about the species scale can be found 
  in~\cite{Castellano:2022bvr,vandeHeisteeg:2023ubh,Cribiori:2023ffn,Blumenhagen:2023yws,Cribiori:2023sch,Calderon-Infante:2023uhz,vandeHeisteeg:2023dlw,Castellano:2023aum,Castellano:2023stg,Castellano:2023jjt,Basile:2024dqq}
and  recently further arguments
 for the agreement of these two definitions of the species scale were provided in~\cite{Bedroya:2024ubj}.}
\begin{equation}
l_{\rm sp}
\equiv \Lambda_{\rm sp}^{-1} \equiv L_{\rm min}. 
\end{equation}
It follows that the species scale is related to the entropy of the minimal black hole in the following way:
\begin{equation}\label{SBHmin}
\Lambda_{\rm sp}={M_p\over (S_{\rm BH,{\rm min}})^{1\over d-2}}\, .
\end{equation}
Employing the BHEDC to minimal black holes one can then infer that the limit of vanishing species scale $ \Lambda_{\rm sp}\rightarrow 0$ is at infinite distance in the space 
 of EFTs, with a distance given as 
 \begin{equation}
d( \Lambda_{\rm sp,i}, \Lambda_{\rm sp,f})=\left|\ln (\Lambda_{\rm sp,f}/\Lambda_{\rm sp,i}) \right|,
\end{equation}
and with a tower of massless states,
in which masses scale as
 \begin{equation}
m \sim( \Lambda_{\rm sp})^{\gamma}\;,
\end{equation}
where $\gamma$ is a positive constant. 
Combing this with the AdS-DC it follows that the cosmological constant $\Lambda$ and the species scale, i.e. the UV cut-off of the EFT, are parametrically related at the boundary of the moduli space as 
 \begin{equation}
| \Lambda|\simeq ( \Lambda_{\rm sp})^{\gamma/\alpha}\, .
 \end{equation}
 So the limit of a small cosmological constant goes along with the limit of a small species scale and vice versa.
 Furthermore minimal black holes and the species scale can also be used to introduce the notion of species thermodynamics~\cite{Cribiori:2023ffn,Basile:2024dqq}, which connects the thermodynamic properties of the KK and string species
 to an entropy and temperature functional over the moduli space of vacua.

The DC, the AdS-DC, and the BHEDC can be further generalized for more general manifolds using geometric flow 
equations~\cite{Kehagias:2019akr,DeBiasio:2020xkv,DeBiasio:2022omq,Velazquez:2022eco,DeBiasio:2022nsd,DeBiasio:2022zuh,DeBiasio:2023hzo}.
Specifically,  the  following  {\bf Ricci flow distance conjecture}  was formulated~\cite{Kehagias:2019akr}:
For a $d$-dimensional Riemannian manifold, the  distance  in the field space of the background metrics along the  Ricci-flow 
is determined by the scalar curvature ${\cal R}(g)$:
\begin{equation}
d({\cal R}_{\rm i}, {\cal R}_{\rm f})=\left|\ln ({\cal R}_{\rm f}/{\cal R}_{\rm i}) \right|.
\end{equation}
Here ${\cal R}_i$ and ${\cal R}_f$ are the corresponding initial and final values of the scalar curvature.
Furthermore it is stated that 
at ${ {\cal R}}=0$ there is an infinite tower of additional massless states in quantum gravity.

One more related distance conjecture, the {\bf large D distance conjecture}~\cite{Bonnefoy:2020uef,DeBiasio:2021yoe},  states that the limit of large space-time dimension, $d\rightarrow\infty$, is also at infinite distance in the space of quantum gravity vacua.

Another interesting limit arises with the gravitino mass going to zero. 
This is because the gravitino mass is generally related to the scale
of spontaneous supersymmetry (SUSY) breaking in non-supersymmetric vacua
(like the one we happen to live in)~\cite{Antoniadis:1988jn,Antoniadis:1990ew}. The {\bf gravitino conjecture}
states~\cite{Cribiori:2021gbf,Castellano:2021yye} that in a supersymmetric theory with a non-vanishing gravitino mass $m_{3/2}$, a tower of states becomes light in the limit $m_{3/2} \to 0$ according to
\begin{equation}
  m \sim \left(\frac{m_{3/2}}{\bar{M}_p}\right)^{\delta} \ \bar{M}_p
\end{equation}
where $\delta$ is an order-one positive parameter and $\bar M_p =
M_p/\sqrt{8 \pi}$ is the reduced Planck mass. 

In closing, we review one last conjecture, which is not linked to
a tower of states, but will be relevant for our discussion. The {\bf weak gravity conjecture} (WGC) states~\cite{Arkani-Hamed:2006emk,Harlow:2022ich} that for given a gauge theory, weakly coupled to Einstein gravity, there exists a charged state with
\begin{equation}
\frac{Q}{m} \geq \left. \frac{{\cal Q}_{\rm BH}}{M_{\rm BH}} \right|_{\rm extremal} = {\cal O}(1) 
\label{WEC}
\end{equation}
in Planck units, where ${\cal Q}_{\rm BH}$ and $M_{\rm BH}$ are the charge and mass of
an extremal black hole, $Q = qg$, $q$ is the quantized charge of the state, and $g$ is the gauge coupling.

The WGC can be seen as a kinematic
requirement that allows extremal black holes to decay. As can be
inferred from the {\it weak cosmic censorship}~\cite{Penrose:1969pc}, charged black
holes must satisfy an extremality bound in order to avoid the presence
of naked singularities. Intuition about the extremality bound
(\ref{WEC}) arises spontaneously from the
 the Reissner-Nordstr\"om (RN) metric that describes
the simplest extremal black hole, which has its mass $M_{\rm BH}$ equal to its
charge ${\cal Q}_{\rm BH}$ in appropriate units. Note that if ${\cal
  Q}_{\rm BH} = M_{\rm BH}$
the single option for the black hole to decay is if there
exists a particle whose charge to mass ratio is at least one.
If $M_{\rm BH} > {\cal Q}_{\rm BH}$ the black hole has inner (Cauchy)
and outer (event) horizons, whereas if $M_{\rm BH}<
{\cal Q}_{\rm BH}$, the RN metric describes a naked singularity.   

\section{Foundations of the Dark Dimension}

\subsection{In which region of the landscape do we live?}

Applying the AdS-DC to dS
(or quasi dS) space provides one possible pathway
to answer this question. Moving forward, we will walk along this pathway.

Many asymptotic limits are expected to have towers of KK
modes. According to the dS-DC, the vacuum energy scales as $\Lambda \sim m^{1/\alpha}$, with $m$ the mass scale of the leading tower~\cite{Lust:2019zwm}. Since the KK
tower contains massive spin-2 bosons, there is a strong constraint from fundamental physics, {\it
  unitarity}, which is expressed in the form of the Higuchi bound and imposes an absolute
  upper limit  $\alpha = 1/2$~\cite{Higuchi:1986py}. Besides, $\alpha$ has a lower limit set by contributions of the Casimir energy; in
  four dimensions $\alpha \geq 1/4$~\cite{Montero:2022prj}. Another property of
asymptotic limits is that both $m$ and $\Lambda$ are very small in
Planck units. Therefore, since the observed amount of dark energy in our world is also
very
small, $\Lambda \sim 10^{-122} M_p^4$, it is
tempting to speculate whether we could be living near an asymptotic
limit. We will then assume that we live within sight of the space boundary in
some infinite distance limit. This assumption automatically leads to the prediction~\cite{Montero:2022prj} of a tower of light fields at the energy scale
\begin{equation}\label{mkklambda}
  m_{\rm KK} \lesssim \Lambda^{1/4} \sim 2.31~{\rm meV} \, .
\end{equation}
Since we have not seen yet experimental evidence of such a tower, it must couple very weakly (if at all) to SM fields.

The ESC connects infinite
distance limits with the
decompactification of $n$ extra dimensions. Now, consistency of large-distance black hole physics in the
presence of a tower of $N$ light fields imposes a bound on the
gravitational cutoff of the EFT, and the fundamental length is no
longer $l_p = M_p^{-1}$, but rather the species length~\cite{Dvali:2007hz,Dvali:2007wp}
\begin{equation}
l_{\rm sp} = \sqrt{N} \  l_p \, .
\label{lN}
\end{equation}
The origin of (\ref{lN}) can be traced back using different
arguments, we follow here the reasoning given in~\cite{Antoniadis:2014bpa} based on quantum
information storage~\cite{Dvali:2008ec}. Consider a pixel of
size $L$ containing $N$ species storing information. The minimal energy required to localize $N$ wave functions
is of order $N/L$. This energy can be associated to a Schwarzschild radius
$r_s = N/ (LM_p^2)$, which must be smaller
than the pixel size if we want to avoid the system to collapse into a
black
hole. Now, $r_s \leq L$, implies there is a minimum size $l_{\rm sp}
\equiv \Lambda_{\rm sp}^{-1} \equiv L_{\rm min} = \sqrt{N} \, M_p^{-1}$ associated to the
scale where gravity becomes strongly coupled and the EFT necessarily
breaks down. Since $l_{\rm sp}$ constitutes the
  smallest black hole size described by the EFT (involving only the
  Einstein term), $\Lambda_{\rm sp}$ codifies the ``number of light degrees
  of freedom'' (i.e., the number of KK excitations lighter than the
  cut-off), given
by $N \sim R_\perp^{n} l_{\rm sp}^{-n}$,
up to energies of order
\begin{equation}
M_* \sim \Lambda_{\rm sp} = m_{\rm KK}^{n/(n+2)} M_p^{2/(2+n)} \,, 
\label{Lambdasp}
\end{equation}
where $n$ is the number of decompactifying  dimensions of radius
$R_\perp \sim
m_{\rm KK}^{-1}$ and where the species scale $\Lambda_{\rm sp}$ corresponds to
 the Planck scale $M_*$ of the higher dimensional
 theory.
 
Decompactification limits are tightly
constrained by observations. Indeed, astrophysical
bounds from the requirement that neutron stars are not excessively heated by KK decays lead to a very restrictive limit on
the mass scale of the KK tower, which depends on the number of
dimensions that are decompactifying: $R_\perp \leq 44~\mu{\rm m}$ for $n=1$, and  $R_\perp \leq
10^{-4}~\mu{\rm m}$ for $n=2$~\cite{Hannestad:2003yd}. The limit on
$R_\perp$ becomes more restrictive with
rising $n$. For $n=1$, deviations from Newton's law impose a more
restrictive constraint, $R_\perp \leq 30~\mu{\rm
  m}$~\cite{Lee:2020zjt}. Remember that the starting point here is the
hypothesis that associates the cosmological constant to a
tower of states whose mass scale satisfies
\begin{equation}
 m_{\rm KK} \simeq  \Lambda^{1/4}/\lambda\,, 
\label{Rperpendicular}
\end{equation}
where 
$\lambda$ is an order one parameter (more on this below).
Now, we have seen that $\Lambda^{-1/4} \sim
85\mu{\rm m}$ and so from (\ref{Rperpendicular}) we conclude that if
$n=2$ the hypothesis is excluded by experiment, but if $n=1$ the
hypothesis can be made compatible with the experiment adjusting the
proportionality factor, which is estimated to be within the range $10^{-4} \lesssim \lambda \lesssim 10^{-2}$. 
Substituting $n=1$ into (\ref{Lambdasp}) we have
$M_* \sim m_{\rm KK}^{1/3} M_p^{2/3}$, and so $10^9 \lesssim M_*/{\rm
  GeV} \lesssim 10^{10}$.\footnote{Auger data of highest energy cosmic rays favor $M_*
  \sim 10^{10}~{\rm GeV}$~\cite{Anchordoqui:2022ejw,Noble:2023mfw}.} 
  We can combine this with the relation between $m_{\rm KK} $ and $\Lambda_{\rm sp}$  to derive the following relation
\begin{equation}
M_* \sim \Lambda_{\rm sp} = \Lambda^{1/12}\ \lambda^{-1/3} \ M_p^{2/3} \,.
\end{equation}

In summary, when astrophysical bounds and gravitational tests of Newton's law are combined with the requirement
that the size of the extra dimension is related to the cosmological
constant, we arrive at the conclusion encapsulated
in (\ref{Rperpendicular}); namely, that there is one extra dimension of
radius $R_\perp$ in the micron range, and that the lower bound for $\alpha =
1/4$ is basically saturated~\cite{Montero:2022prj}. Because of its connection to the observed dark energy, this
dimension has been nicknamed {\it the dark dimension}.

In closing we note that explicit string calculations of the vacuum energy~(see
e.g.~\cite{Itoyama:1986ei,Itoyama:1987rc,Antoniadis:1991kh,Bonnefoy:2018tcp})
show that the lower limit on $\alpha$ is saturated. In particular, a
strongly warped throat with its redshifted KK tower provides a natural
string theoretic mechanism that realizes the scaling $m \sim \Lambda^{1/4}$, with the dark dimension being the one along the throat~\cite{Blumenhagen:2022zzw}. Actually, a theoretical
amendment on the connection between the cosmological and KK mass scales confirms $\alpha =
1/4$~\cite{Anchordoqui:2023laz}.\footnote{Unsubstantiated criticisms raised  in~\cite{Branchina:2023ogv,Branchina:2024ljd} have
  been addressed in~\cite{Anchordoqui:2023laz}. We reiterate herein that quantum gravity and string theory are different from field theory and lead to a finite
result relating $\Lambda$ to the KK scale via (\ref{Rperpendicular}). Two arguments support this statement:
{\it (i)}~the AdS-dC stressing that all infinities cancel out (for
AdS, this conjecture holds in many examples) and {\it (ii)}~the
explicit one-loop string computation, which is finite due to modular
invariance and confirms $\alpha =1/4$.} 

\subsection{The scale of SUSY breaking}

It is of course interesting to explore whether there is a relation between the SUSY breaking scale and
the measured value of the dark energy density
$\Lambda$. Such a relation can be derived 
by combining two quantum gravity consistency swampland constraints,
which tie $\Lambda$ and the gravitino mass $m_{3/2}$, to the mass
scale of a light KK tower and, therefore, to the UV cut-off of the
EFT~\cite{Antoniadis:1988jn,Cribiori:2021gbf,Castellano:2021yye}. One 
can then use the constraint
on $m_{3/2}$ to infer the implications of the dark dimension scenario for the
scale of supersymmetry breaking. In general, one can distinguish two
situations. In the first case, the gravitino mass and the cosmological
constant are related to the same tower of states. This is arguably the
simplest scenario, in which the natural scale for
SUSY signatures is of order $\Lambda^{1/8} \sim {\rm TeV}$, and
therefore is
within reach of LHC and/or of the next generation of hadron
colliders~\cite{Anchordoqui:2023oqm}.
In the second case, $m_{3/2}$ and $\Lambda$ are related to different
towers. This scenario requires a decoupling of the gravitino mass from the cosmological constant and is thus more difficult to realize in concrete models.

Possible string theory and effective supergravity
realizations of the dark dimension scenario with broken supersymmetry
are discussed in~\cite{Anchordoqui:2023oqm}. 

\subsection{The dark dimension as a space with two boundaries}

It was recently conjectured that the dark dimension can be viewed
as a line interval with end-of-the-world 9-branes (EW9-brane) attached
at each end~\cite{Schwarz:2024tet}. This construction derives from the 10D $E_8 \times E_8$
heterotic string theory in the strong coupling limit that has an orbifold $S^1/\mathbb{Z}_2$
eleventh dimension (the dark dimension), which is not visible in
perturbation theory. The gravitational field propagates in the bulk
$\mathbb{R}^{10} \times S^1/\mathbb{Z}_2$, whereas the $E_8
\times E_8$ fields propagate only at the $\mathbb{Z}_2$ fixed points on the two EW9-branes.

The 10D coupling of the $E_8$ is a fixed number in 11D supergravity (M-theory) units. When six out of ten space-time
dimensions are compactified on a Calabi-Yau (CY6) manifold, the tree-level
couplings of the effective 4D theory are simply given by the
volume $V$ of CY6 (always in $M_{11}$
units)~\cite{Witten:1996mz,Banks:1996ss}. The SM gauge couplings $(g)$ are observed to be
${\cal O} (1)$, and this forces $V$ on one of the branes to be ${\cal O}(1)$ in $M_{11}$ units.

For smooth compactifications, the volume of CY6 at given position of the coordinate $x^{11}$,
\begin{equation}
v(x^{11}) = \int_{{\rm CY}6} \sqrt{g} \ d^6x \,,
\end{equation}
becomes an approximate linear function of the
extra coordinate, and decreases from one $E_8$ to the other
$E_8$~\cite{Witten:1996mz}. This implies that $v$ has different values at the two fixed
points $x^{11} = 0$ and $x^{11} = \pi \rho$. Herein we identify $V = v(0)$
and $V' = v(\pi \rho)$. Note that when the theory in one $E_8$ is perturbative the theory
on the other $E_8$ becomes non-perturbative when the radius $\rho$
of the dark dimension is
large. This forces an upper bound on the size of the 11th dimension to avoid $V'$ to become negative. If the anomaly
coefficient is ${\cal O} (1)$, then $\rho \sim 1~\mu{\rm m}$ is not
allowed. This is a general phenomenon in orientifold compactifications in which
couplings of localized interactions acquire a linear dependence in
the extra dimension if one extra dimension (transverse to the brane) is large~\cite{Antoniadis:1998ax}. The absence of
such divergences requires local tadpole cancellation (between branes and orientifolds).

However, there are some particular non-geometric compactifications
where the correction to the other $E_8$ coupling vanishes and there is
no constraint on $\rho$~\cite{Caceres:1996is}. In such a particular case, we have the
following connections between the 11D Planck mass, $M_{11}$ (defined in terms of
the coefficient of the Einstein Lagrangian in 11D supergravity, as $M_{11} = \kappa^{-2/9}$), 
the  radius of the dark dimension $\rho$, and the compactification
radius $R = V^{1/6}$,
\begin{equation}
  \kappa = (2 \alpha)^{3/4} R^{9/2} \,,
\end{equation}
and
\begin{equation}
  \rho = (\alpha/2)^{3/2} \ M_p^2 \ R^3 \, 
\end{equation}
where $V$ is the CY6 volume on the SM boundary and $\alpha \equiv g^2/(4\pi)$. 
For $\rho \sim 1~\mu{\rm m}$ and phenomenological value $\alpha \sim
1/25$, we obtain $M_{\rm KK} \sim R^{-1} \sim 7 \times 10^8{\rm
  GeV}$ and $M_{11} \sim 10^9{\rm GeV}$.

In summary, when six dimensions ($x^5,\cdots, x^{10}$) are
compactified on a CY manifold, the eleven-dimensional bulk of the
world becomes 5D while the 9-branes at its boundaries become
3-branes. The entire SM lives on one of those 3-branes and is
oblivious to the bulk of the 5D world or its other boundary. The
threshold structure of the $\rho \gg R \gtrsim l_{\rm sp} \sim M_{11}^{-1}$ regime of
the M-theory can be summarized as follows:
\begin{itemize}[noitemsep,topsep=0pt]
\item Gravity has a threshold at a rather low energy scale $m_{\rm KK}
  \sim 1/\rho \sim {\rm eV}$  above which it becomes 5D. However, this
  threshold does not affect any gauge, Yukawa or scalar forces of the
  SM, which remains 4D at distances shorter than $\rho$.
\item The next threshold arises at the KK scale $M_{\rm KK} \sim 1/R
  \sim 7 \times 10^8~{\rm GeV}$, where six more dimensions open up for both gravity and gauge interactions.   
\item Almost immediately above this scale (around $\Lambda_{\rm sp}$), the effective field theory description breaks down and the fully quantized M-theory (whatever that is) takes over.  
\end{itemize}

\section{Dark Matter Candidates}

The dark dimension provides a colosseum for dark matter contenders. In
this section we review the general properties of the various dark matter candidates.

\subsection{Primordial black holes}
\label{sec:4i}

It has long been speculated that black holes could be produced from the
collapse of large amplitude fluctuations in the early
universe~\cite{Zeldovich:1967lct,Hawking:1971ei,Carr:1974nx,Carr:1975qj}. For
an order of magnitude estimate of the black hole mass $M_{\rm BH}$, we first
note that the cosmological energy density scales with time $t$
as $\rho \sim 1/(Gt^2)$ and the density  needed for a region of mass
$M_{\rm BH}$ to collapse within its Schwarzschild radius is $\rho \sim
c^6/(G^3M_{\rm BH}^2)$, so that primordial black holes (PBHs) would initially
have around the cosmological horizon mass~\cite{Carr:2020xqk}
\begin{equation}
M_{\rm BH} \sim \frac{c^3 t}{G} \sim 10^{15}
\left(\frac{t}{10^{-23}~{\rm s}}\right)~{\rm g} \, .
\end{equation}
This means that a black hole would have the reduced Planck mass ($\bar M_p
\sim 10^{-5}~{\rm g}$) if they formed at the Planck time
($10^{-43}~{\rm s}$), $1~M_\odot$ if they formed at the QCD epoch
($10^{-5}~{\rm s}$), and $10^{5} M_\odot$ if they formed at $t \sim
1~{\rm s}$, comparable to the mass of the holes thought to reside in galactic nuclei. This back-of-the-envelope calculation suggests that PBHs could span an enormous mass
range. Despite the fact that the mass spectrum of these PBHs is yet to
be shaped, on cosmological scales they would behave like a typical cold dark
matter particle.

However, an all-dark-matter interpretation in terms of PBHs is severely constrained by
observations~\cite{Carr:2020xqk,Green:2020jor,
  Villanueva-Domingo:2021spv,LISACosmologyWorkingGroup:2023njw}.  To be specific, the extragalactic
$\gamma$-ray background~\cite{Carr:2009jm}, the cosmic microwave
background (CMB)~\cite{Clark:2016nst}, the 511~keV $\gamma$-ray
line~\cite{DeRocco:2019fjq,Laha:2019ssq,Dasgupta:2019cae,Keith:2021guq},
EDGES 21-cm signal~\cite{Mittal:2021egv}, and the MeV Galactic diffuse emission~\cite{Laha:2020ivk,Berteaud:2022tws,Korwar:2023kpy} constrain evaporation of black holes with masses $\lesssim 10^{17}~{\rm g}$,
whereas the non-observation of microlensing events by MACHO~\cite{Macho:2000nvd}, EROS~\cite{EROS-2:2006ryy},
  Kepler~\cite{Griest:2013aaa}, Icarus~\cite{Oguri:2017ock},
  OGLE~\cite{Niikura:2019kqi} and Subaru-HSC~\cite{Croon:2020ouk} set
  an upper limit on the black hole abundance for masses $M_{\rm BH} \gtrsim
  10^{21}~{\rm g}$.

Before proceeding, we pause and call attention to a captivating
coincidence:
\begin{equation}
{\tt size \ of \ the  \ dark \ dimension
}  \sim {\tt wavelength \ of \ visible \ light} \,,
\label{stun}
\end{equation}
which implies that the
  Schwarzschild radius of 5D black holes is well below the wavelength
  of light. For point-like lenses, this is precisely the critical length where
  geometric optics breaks down and the effects of wave optics suppress
  the magnification, obstructing the sensitivity to 5D PBH
  microlensing signals~\cite{Croon:2020ouk}.
  So 5D PBHs escape these microlensing constraints; at the same time, as pointed out in~\cite{Anchordoqui:2022txe},
  they are: bigger, colder, and longer-lived than a
usual 4D black hole of the same
mass.

Throughout we rely on the probe brane approximation, which ensures that the
only effect of the brane field is to bind the black hole to the
brane~\cite{Giddings:2001bu}. This is an adequate approximation provided $M_{\rm BH}$ is well
above the brane tension, which is presumably of the order of but
smaller than $M_\star$. We also assume that the
black hole can be treated as a flat $d$ dimensional object. This
assumption is valid for extra dimensions that are larger than the
5D Schwarzschild radius, which is given by~\cite{Tangherlini:1963bw, Myers:1986un,Argyres:1998qn} 
  \begin{equation}
    r_s(M_{\rm BH}) \sim \frac{1}{M_*} \left[ \frac{2}{3\, \pi}
      \ \frac{M_{\rm BH}}{M_*}
    \right]^{1/2} \, .
\end{equation}

Schwarzschild black holes radiate all particle species lighter than or comparable to
their Hawking temperature, which in four dimensions is related to the mass of the black hole by
\begin{equation} 
T_H = \frac{\bar M_p^2}{8 \pi M_{\rm BH}} \sim 
\, \bigg(\frac{M_{\rm BH}}{10^{16} \, {\rm g}} \bigg)^{-1}~{\rm MeV}\,,
\label{T4d}
\end{equation}
whereas for 5D black holes the Hawking temperature mass relation
is found to be~\cite{Anchordoqui:2024dxu}
\begin{equation}
  T_H  \sim \frac{1}{r_s} \sim  
    \left(\frac{M_{\rm BH}}{10^{12}~{\rm g}}\right)^{-1/2}~{\rm MeV} \, .
\label{tempes}
  \end{equation}
The numerical estimate of (\ref{tempes})  applies to
the dark dimension scenario with $M_* \sim 10^{10}~{\rm GeV}$.\footnote{We have taken the
highest possible value of $M_*$ to remain conservative in the
estimated bound on the fraction of dark matter composed of primordial black holes  $f_{\rm PBH}$.}  It is evident that 5D
black holes are colder than 4D black holes of the same mass.

Armed with the Hawking temperature, we can now calculate the entropy
of the 5D black hole~\cite{Anchordoqui:2001cg}
\begin{equation}
  S_{\rm BH} = \frac{4}{3} \pi \ M_{\rm BH} \ r_s \, .
\end{equation}
In the rest frame of the Schwarzschild black hole, both the average
number~\cite{Hawking:1974rv,Hawking:1975vcx} and the probability
distribution of the number~\cite{Parker:1975jm,Wald:1975kc,Hawking:1976ra} of outgoing particles in each mode
obey a thermal spectrum. However, in the neighborhood of the horizon
the black hole produces an effective potential barrier that
backscatters part of the emitted radiation, modifying the thermal
spectrum. The so-called ``greybody factor'', which controls
the black hole absorption cross section, depends upon the spin of the
emitted particles $s$, their energy $Q$, and $M_{\rm BH}$~\cite{Page:1976df,Page:1976ki,Page:1977um,Kanti:2002nr,Ireland:2023zrd}. The prevailing energies of the emitted particles are
$\sim T_{\rm H} \sim 1/r_s$, resulting in $s$-wave dominance of the
final state.  This implies that the black hole evaporates with equal
probability to a particle on the brane and in the compact
space~\cite{Emparan:2000rs,Dimopoulos:2001hw}. Thereby, the process of evaporation is
driven by the large number of SM brane modes.

The
Hawking radiation causes a 4D black hole to lose mass at the following rate~\cite{Keith:2021guq}
\begin{eqnarray}
 \left. \frac{dM_{\rm BH}}{dt}\right|_{\rm evap} & = & -\frac{\bar
                                                       M_p^2}{30720 \
                                                       \pi \ M_{\rm BH}^2} \ \sum_{i} c_i(T_H) \ \tilde f \ \Gamma_s
    \nonumber \\ 
   & \sim & -7.5 \times 10^{-8} \ \left(\frac{M_{\rm BH}}{10^{16}~{\rm
                                                   g}}\right)^{-2} \ \sum_{i}
      c_i (T_H)  \ \tilde f
 \ \Gamma_s~{\rm g/s}   \,, 
\label{cinco}
\end{eqnarray}
whereas a 5D black hole has an evaporation rate of~\cite{Anchordoqui:2024dxu}
\begin{eqnarray}
  \left. \frac{dM_{\rm BH}}{dt}\right|_{\rm evap}
& \sim & - 9 \ \pi^{5/4} \zeta(4) T_H^2 \ \sum_{i} c_i(T_H) \ \tilde f
      \ \Gamma_s 
    \nonumber \\
& \sim & - 1.3 \times 10^{-12} \ \left(\frac{M_{\rm BH}}{10^{16}~{\rm g}}\right)^{-1} \ \sum_{i}
      c_i (T_H) \ \tilde f
\  \Gamma_s~{\rm g/s}  \,,     
\label{seis}
\end{eqnarray}
where $c_i(T_H)$ counts the number of internal degrees of freedom of particle 
species $i$ of mass $m_i$ satisfying $m_i \ll T_H$,  $\tilde f = 1$  $(\tilde f = 7/8)$ for bosons
(fermions), and where $\Gamma_{s=1/2} \approx 2/3$ and $\Gamma_{s=1} \approx
1/4$ are the (spin-weighted) dimensionless greybody factors normalized to the black
hole surface area~\cite{Anchordoqui:2002cp}.
Now, comparing (\ref{cinco}) and (\ref{seis}) it is easily seen that 5D black holes live
longer than 4D black holes of the same mass.

Integrating (\ref{seis}) we can parametrize the 5D black hole
lifetime as a function of its mass and temperature,
\begin{equation}
  \tau_s \sim 13.8 \ \Bigg(\frac{M_{\rm BH}}{10^{12}~{\rm g}} \Bigg)^2 \ \left(\frac{6}{\sum_{i}
      c_i (T_s) \ \tilde f
      \  \Gamma_s}\right)~{\rm Gyr} \, ,
\label{lifetime}  
\end{equation}
where we have used (\ref{tempes}) to estimate that
$T_H \sim 1~{\rm MeV}$ and therefore $c_i(T_H)$ receives a
contribution of 6 from neutrinos, 4 for electrons, and 2 from photons,
yielding $\sum_{i} c_i (T_H) \ \tilde f \ \Gamma_s = 6$.  Armed with
(\ref{lifetime}) we can estimate the bound on the 5D PBH abundance by
a simple rescaling procedure of the $d=4$ bounds on the fraction of
dark matter composed of primordial black holes $f_{\rm PBH}$. The key
point for such a rescaling is that for a given photon energy, or
equivalently a given Hawking temperature, we expect a comparable limit
on $f_{\rm PBH}$ for both $d=4$ and $d=5$. For example, from
(\ref{T4d}) and (\ref{tempes}) we see that the constraint of
$f_{\rm PBH} \lesssim 5 \times 10^{-5}$ for 4D black holes with
$M_{\rm BH} \sim 10^{16}~{\rm g}$~\cite{Korwar:2023kpy}, should be roughly the same for the
abundance of 5D black holes with $M_{\rm BH} \sim 10^{12}~{\rm
  g}$. Now, since in $d=4$ for $M_{\rm BH} \sim 4 \times 10^{17}~{\rm g}$ we
have $f_{\rm PBH} \sim 1$~\cite{Korwar:2023kpy}, this implies the same abundance for 5D
black holes of $M_{\rm BH} \sim 10^{15}~{\rm g}$. By duplicating this
procedure for heavier black holes we conclude that for a species scale
of ${\cal O} (10^{10}~{\rm GeV})$, an all-dark-matter interpretation in
terms of 5D black holes must be feasible for masses in the range~\cite{Anchordoqui:2024dxu}
\begin{equation}
10^{15} \lesssim M_{\rm BH} /{\rm
  g} \lesssim 10^{21} \, .
\label{massrange}
\end{equation}
This range is extended compared to that in the
4D theory by more than two orders of magnitude in the low mass region.

Extremal black holes trace the boundary between black-hole
configurations and horizonless naked singularities. As put forward by
the WGC, they are
characterized by the minimally allowed mass (radius) for a given
amount of black-hole charge ${\cal Q}_{\rm BH}$ (or angular
momentum). Near-extremal black holes are characterized by a finite
(non-zero) 
mass gap of the first excited state above the extremal
(zero-temperature) black-hole configuration~\cite{Maldacena:1996ds}. The temperature of such
near-extremal black holes is found to be
\begin{equation}
  T_{ne} \sim \frac{\beta^{1/2} T_H}{S_{\rm BH}^{1/2}} \,,
\end{equation}  
where $\beta$ is a factor of order-one that controls the differences
between $M_{\rm BH}$ and ${\cal Q}_{\rm BH}$; for details
see~\cite{Cribiori:2022cho,Basile:2024dqq}.
If there were 5D primordial near-extremal black holes in nature, then an all-dark-matter interpretation would be possible in the mass range~\cite{Anchordoqui:2024akj}
\begin{equation}
10^{5} \sqrt{\beta} \lesssim M_{\rm BH}/{\rm
  g} \lesssim 10^{21} \, .
\label{massrange2}
\end{equation}

Up until now we have assumed that the 5D black holes stay attached to the
brane during the evaporation process. Hereafter we relax this
assumption and allow them wander off into the bulk. Without knowing
more details of the bulk and brane theory it is not worth considering to calculate the probability of
such wandering in detail. However, we can assume that the black holes are
out of the brane-world and study the evaporation effects of these bulk
PBHs. Furthermore, it is always possible that the PBHs are
produced in the bulk to start with. This situation will be more
appealing within the model discussed in Sec.~\ref{sec:6}, in which we theorize that the dark dimension
may have undergone a uniform rapid expansion, together with the
three-dimensional non-compact space, by regular exponential inflation
driven by an (approximate) higher dimensional cosmological constant. If this were the case, then primordial fluctuations during inflation of the compact space could lead to the
production of black holes in the bulk. In what follows, we then assume
that PBHs are
localized or propagate in the bulk.
 
Bulk black holes live longer than those attached to the brane. This
is because KK modes are excitations in the full transverse space and so their
overlap with small (higher dimensional) black holes is suppressed by the
geometric factor $(r_s/R_\perp)$ relative to the brane fields. This
geometric suppression precisely compensates for the enormous number of
modes and the total KK contribution is only of same order as that from
a single brane field~\cite{Emparan:2000rs}. Actually, greybody factors suppress graviton
emission when compared to fermions and gauge bosons, and hence bulk
black holes which do not have access to the brane degrees of freedom
are expected to live longer. In
addition, since there is no emission on the brane the bounds due to
photon evaporation can be avoided. This implies that PBHs localized in
the bulk can provide an all-dark-matter interpretation if
\begin{equation}
  10^{11} \lesssim M_{\rm
    BH} /{\rm g} < 10^{21} \,,
\label{cuarentaycuatro}
\end{equation}
where we have remained conservative, and following~\cite{Han:2002yy} we assumed that the ratio of the emitted flux into a single
brane field over a single bulk field is
 roughly a factor of two~\cite{Anchordoqui:2024dxu}.

\subsection{KK gravitons}
\label{sec:4ii}

It was observed in~\cite{Gonzalo:2022jac} that
the universal coupling of the SM fields to the massive
spin-2 KK excitations of the graviton in the dark dimension provides
an alternative dark matter candidate. Within this model the cosmic evolution of the hidden sector is primarily dominated by ``dark-to-dark'' decays, yielding
  a specific realization of the dynamical dark matter
  framework~\cite{Dienes:2011ja}. Consider a tower of equally spaced dark gravitons,
indexed by an integer $l$, and with mass  $m_l = l \ m_{\rm KK}$. The partial decay width of KK graviton $l$ to SM fields is found to be,
\begin{equation}
  \Gamma^l_{{\rm SM}} = \frac{\tilde{\lambda}^2 \ m^3_{\rm KK} \ l^3}{80 \pi
    \bar M_p^2} \,,
\label{Ggamma}
\end{equation}
where $\tilde \lambda$ takes into account all the available decay channels and is a function of
time~\cite{Hall:1999mk}.

In the absence of isometries
in the dark dimension, which is the common expectation, the KK momentum of the
dark tower is not conserved. This means that a dark graviton of KK
quantum $n$ can decay to two other ones, with quantum numbers $n_1$
and $n_2$. If the KK quantum violation can go
up to $\delta n$, the number of available channels is roughly $l \,
\delta n$. In addition, because the decay is almost at threshold, the phase space
factor is roughly the velocity of decay products,
  $v_{\rm r.m.s.} \sim \sqrt{m_{\rm KK} \ \delta n /m_l}$. Putting all
  this together we obtain the total
  decay width, 
\begin{eqnarray}
  \Gamma^l_{\rm tot} & \sim & \sum_{l'<l} \ \ \sum_{0<l''<l-l'} \Gamma^l_{l' l''} \sim
  \beta^2 \frac{m_l^3}{\bar M_p^2} \times \frac{m_l}{m_{\rm KK}} \ \delta
                              n \times \sqrt{\frac{m_{\rm KK} \delta_n}{m_l} } \nonumber \\
  & \sim & \beta^2 \
    \delta n^{3/2} \frac{m_l^{7/2}}{\bar M_p^2 m_{\rm KK}^{1/2}} \,,
 \end{eqnarray}   
where $\beta$ parametrizes our ignorance of decays in the dark dimension~\cite{Gonzalo:2022jac}.

To estimate the time
evolution of the dark matter mass 
assume that for times larger than $1/\Gamma^l_{\rm tot}$ dark matter
which is
heavier than the corresponding $m_l$ has already decayed, and so it
follows that
\begin{equation}
  m_l \sim \left(\frac{\bar M_p^4 \ m_{\rm KK}}{\beta^4 \ \delta n^3}\right)^{1/7} t^{-2/7} \,,
\label{mt}
\end{equation}
where $t$ indicates the time elapsed since the big bang~\cite{Gonzalo:2022jac}.

Consistency with CMB anisotropies requires  $\Gamma^l_{\gamma \gamma} < 5 \times 
10^{-25}~{\rm s}^{-1}$ between the last scattering surface and
reionization~\cite{Slatyer:2016qyl}. Taking $\tilde{\lambda} = 1$ (to
set out the decay into photons) and using (\ref{Ggamma}) it follows that the
CMB requirement is satisfied for
$l \lesssim 10^8$  at the time $t_{\rm MR} \sim 6 \times
10^4~{\rm yr}$ of matter-radiation equality. In other words, by setting
$\tilde \lambda \sim 1$ and $m_l (t_{\rm MR}) \lesssim  1~{\rm MeV}$, the evolution of $m_l$ with cosmic time given in (\ref{mt}) is such
that at the last scattering surface the dominant KK state in the
dynamical dark matter ensemble has the correct decay width to
accommodate the CMB constraints~\cite{Law-Smith:2023czn}.

Now, we
have seen that dark matter decay gives the daughter particles a
velocity kick. Self-gravitating dark-matter halos that have a virial
velocity smaller than this velocity kick may be disrupted by these
particle decays. Consistency with existing data requires roughly
$\delta n \sim 1$, and $\beta \sim 635$~\cite{Obied:2023clp,Vafa:2024fpx}. For selected fiducial parameters, the cosmic evolution of the
incredible bulk predicts via (\ref{mt}) a dominant particle mass of $\sim 900~{\rm
  keV}$ at CMB, of $\sim 500~{\rm keV}$ in the Dark Ages, of $\sim
150~{\rm keV}$ at Cosmic Dawn, and of $\sim 50~{\rm keV}$
in the local universe. This is in sharp contrast to typical dark
matter decay scenarios with one unstable particle (such as sterile
neutrinos~\cite{Abazajian:2017tcc}). Simultaneous observations of
signals at Cosmic Dawn and
in the local universe could constitute the smoking gun of the
incredible bulk~\cite{Anchordoqui:2022svl}.

For many purposes, a black hole can be replaced by a bound state of
gravitons~\cite{Dvali:2011aa}. As a matter of fact, a
  correspondence between 5D PBHs and massive KK gravitons as dark matter
  candidates has been conjectured in~\cite{Anchordoqui:2022tgp}.

\subsection{A fuzzy radion}
  
The radion stabilizing the dark dimension could be yet another dark
matter contender~\cite{Anchordoqui:2023tln}. This is because in
principle the
radion could be ultralight, and if this were the case it would serve as a fuzzy dark matter candidate. A simple
cosmological production mechanism brings into play unstable
KK graviton towers which are fueled by the decay of the
inflaton. As in the previous model, the cosmic evolution of the dark sector is
mostly driven by ``dark-to-dark'' decay processes that regulate the
decay of KK gravitons within the dark tower, conveying another
realization of the dynamical dark matter framework~\cite{Dienes:2011ja}. In the spirit
of~\cite{Mohapatra:2003ah}, within this model it is assumed that the intra-KK decays in the bulk carry a spontaneous breakdown
of the translational invariance in the compact space, such that the 5D
momenta are not conserved (but now $\delta n \gg 1$). Armed with these two reasonable assumptions
it is straightforward to see that the energy the inflaton deposited in the KK tower
should have 
collapsed all into the radion well before BBN.  

\section{Neutrino Masses and Mixing}

The dark dimension scenario provides a profitable arena to realize an
old idea for explaining the smallness of neutrino masses by introducing the
right-handed neutrinos as 5D bulk states with Yukawa couplings to the
left-handed lepton and Higgs doublets that are localized states on the
SM brane stack~\cite{Dienes:1998sb, Arkani-Hamed:1998wuz,
  Dvali:1999cn}. The neutrino masses are then suppressed due to the
wave function of the bulk states.

More indicatively, we introduce three 5D Dirac fermions  $\Psi_\alpha$, which are singlets under the SM gauge symmetries and interact in our brane with the three active left-handed neutrinos in a way that conserves lepton number.  The  $S^1/\mathbb{Z}_2$ symmetry contains $x^{11}$ to $-x^{11}$ which acts as chirality ($\gamma_5$) on spinors. In the Weyl basis each Dirac field can be decomposed into two two-component spinors $\Psi_\alpha \equiv (\psi_{\alpha L},\psi_{\alpha R})^T$. 

The generation of neutrino masses
originates in 5D bulk-brane interactions of the form
\begin{equation} 
  \mathscr{L}  \supset h_{ij} \ \overline L_i \ \tilde{H} \ \psi_{jR}(x^{11}=0) \,,
\end{equation}
where $\tilde{H} = -i\sigma_{2}H^{*}$, $L_i$ denotes the lepton
doublets (localized on the SM brane), $\psi_{jR}$ stands for the 3 bulk (right-handed) $R$-neutrinos
evaluated at the position of the SM brane, $x^{11}=0$ in the
dark-dimension coordinate $x^{11}$, and $h_{ij}$ are coupling
constants. This gives a coupling with the
$L$-neutrinos of the form $\langle H \rangle \  \overline{\nu}_{L_i} \
\psi_{jR} (x^{11}=0)$, where $\langle H \rangle = 175~{\rm GeV}$ is the Higgs vacuum
expectation value. Expanding $\psi_{jR}$ into modes canonically normalized leads for each of them to a Yukawa $3 \times 3$ matrix suppressed by the square root of the volume of the bulk
$\sqrt{\pi R_\perp M_s}$, i.e.,
\begin{equation}
Y_{ij}= \frac{h_{ij}}{\sqrt{\pi R_\perp M_s}} \sim h_{ij} \frac{M_s}{M_p} \,,
\end{equation}
where $M_s \lesssim M_*$ is the string scale, and where
in the second rendition we have dropped factors of $\pi$'s and of the string coupling.

Now, neutrino oscillation data can be well-fitted in terms of two
nonzero differences $\Delta m^2_{ij} = m^2_i - m^2_j$ between the
squares of the masses of the three mass eigenstates; namely,
$\Delta m_{21}^2 =(7.53 \pm 0.18) \times 10^{-5}~{\rm eV}^2$ and
$\Delta m^2_{32} = (2.453 \pm 0.033) \times 10^{-3}~{\rm eV}^2$ or
$\Delta m^2_{32} = -(2.536 \pm 0.034) \times 10^{-3}~{\rm
  eV}^2$~\cite{ParticleDataGroup:2022pth}. It is easily seen that to
obtain the correct order of magnitude of neutrino masses the coupling
$h_{ij}$ should be of order $10^{-4}$ to $10^{-5}$ for
$10^9 \lesssim M_s/{\rm GeV} \lesssim 10^{10}$.

Note that KK modes of the
5D $R$-neutrino fields behave as an infinite tower
of sterile neutrinos, with masses proportional to $m_{\rm
    KK}$. However, only the lower mass states of the tower mix with
the active SM neutrinos in a pertinent fashion. The non-observation of
neutrino disappearance from oscillations into sterile neutrinos at
long- and short-baseline experiments places a 90\% CL upper limit  $R_\perp < 0.4~\mu{\rm m}$ for the
normal neutrino ordering, and $R_\perp < 0.2~\mu{\rm m}$ for the
inverted neutrino
ordering~\cite{Machado:2011jt,Forero:2022skg}.\footnote{We arrived at
  these upper bounds by looking at the low mass limit of the lightest
  neutrino state in Fig.~6 of~\cite{Forero:2022skg} and rounding the numbers to one
  significant figure.} This set of parameters corresponds to $\lambda \lesssim 10^{-3}$ and so
$m_{\rm KK} \gtrsim 2.5~{\rm eV}$~\cite{Anchordoqui:2022svl}.

Before proceeding, it is important to stress that the upper bounds
 on $R_\perp$ discussed in the previous paragraph are sensitive to assumptions of the $5^{\rm
  th}$ dimension geometry. Moreover,  in the presence of bulk masses~\cite{Lukas:2000wn,Lukas:2000rg},
the mixing of the first KK modes to active neutrinos can be
suppressed, and therefore the aforementioned bounds on $R_\perp$ can
be avoided~\cite{Carena:2017qhd,Anchordoqui:2023wkm}. It is also worth mentioning that such bulk masses have
the potential to increase the relative importance of the higher KK
modes, yielding distinct oscillation signatures via neutrino
disappearance/appearance effects.

Non-minimal extensions of the dark dimension, in which $m_{3/2}$ and
$\Lambda$ have different KK towers, allow a high-scale SUSY breaking and can therefore host a
rather heavy gravitino together with a modulino with a mass of about
50~eV~\cite{Anchordoqui:2023qxv}. For a  particular example, we note
that the modulino could be the fermionic partner of the radion.\footnote{In the standard moduli stabilization by fluxes,
  all complex structure moduli and the dilaton are stabilized in a
  supersymmetric way while K\"ahler class moduli need an input from SUSY
  breaking. The radion is K\"ahler class and exists in a model 
  independent fashion within the dark
  dimension scenario.} These models with high-scale SUSY
breaking are fully predictive through neutrino-modulino oscillations~\cite{Benakli:1997iu} which
can be confronted with data to be collected by experiments at the
Forward Physics Facility~\cite{Anchordoqui:2021ghd,Feng:2022inv}.

A seemingly different, but in fact closely related subject is the 
the {\it sharpened} version of the WGC forbidding the
presence of non-SUSY AdS vacua supported by fluxes in a consistent
quantum gravity theory~\cite{Ooguri:2016pdq}. This is because (unless
the gravitino is very light, with mass in the meV range) {\it neutrinos have to be Dirac with
right-handed states propagating in the bulk so that the KK neutrino towers compensate for the
graviton tower to maintain stable dS vacua}~\cite{Anchordoqui:2023wkm}.

\section{Mesoscopic Extra Dimension from 5D Inflation}
\label{sec:6}

It is unnatural to entertain that the size of the dark dimension would
remain fixed during the evolution of the Universe right at the species
scale, since the Higuchi bound implies a very low inflation scale. One possible mechanism to accommodate this hierarchy is to
inflate the size of the dark dimension. The required inflationary phase can be described by a 5D dS (or approximate) solution of
Einstein equations, with cosmological constant and a 5D Planck scale
$M_* \sim 10^9~{\rm GeV}$~\cite{Anchordoqui:2022svl}. All dimensions (compact and non-compact)
expand exponentially in terms of the 5D proper time. It is
straightforward to see that this set-up requires about 42 e-folds to
expand the 5th dimension from the fundamental length ${\cal O}
(M_*^{-1})$ to the micron size ${\cal O}
(R_\perp)$. At the end of 5D inflation, or at any given moment, one
can interpret the solution in terms of 4D fields using 4D Planck units
from the relation $M_p^2=M_*^3 R$, which amounts going to the 4D
Einstein frame. This implies that if $M_*^{-1}\leq R \leq R_\perp$ expands $N$ e-folds, then the 3D space would expand $3N/2$ e-folds as a result of a uniform 5D inflation. Altogether, the 3D space has expanded by about 60 e-folds to solve the horizon problem, while connecting this particular solution to the generation of large size extra dimension.

Besides solving the horizon problem, 4D slow-roll inflation predicts
an approximate scale-invariant Harrison-Zel'dovich power spectrum of primordial density
perturbations~\cite{Harrison:1969fb, Zeldovich:1972zz} consistent with
CMB observations~\cite{Planck:2018vyg}. This is due to the fact that the 2-point function of a massless
minimally coupled scalar field in dS space behaves logarithmically at
distances larger than the cosmological horizon, a property which is
though valid for any spacetime
dimensionality~\cite{Ratra:1984yq}. When some dimensions are however
compact, this behaviour is expected to hold for distances smaller than
the compactification length, while deviating from scale invariance at
larger distances, potentially conflicting with observations at large
angles. Remarkably, consistency of 5D inflation with CMB observations is maintained if the
size of the dark dimension is larger than about a micron, implying a
change of behaviour in the power spectrum at angles larger than 10
degrees, corresponding to multiple moments $l\lesssim 30$, where
experimental errors are getting
large~\cite{Anchordoqui:2023etp}. Actually, the scale invariance of the power
spectrum is obtained upon
summation over the contribution of the inflaton KK-modes' fluctuations
that correspond to a tower of scalars from the 4D point of
view. Spectral indices dependence on slow-roll parameters and tensor perturbations have been computed
in~\cite{Antoniadis:2023sya}. The
tensor-to-scalar ratio is found to be $r = 24 \epsilon_V$, and so the
 95\% CL upper limit $r < 0.032$ (derived using a combination of
BICEP/Keck 2018 and {\it Planck} data)~\cite{BICEP:2021xfz,Tristram:2021tvh} places an experimental
constraint on the potential slow-roll parameter: 
$\epsilon_V < 0.0013$.

\begin{figure}[htb!]
  \begin{minipage}[t]{0.49\textwidth}
    \postscript{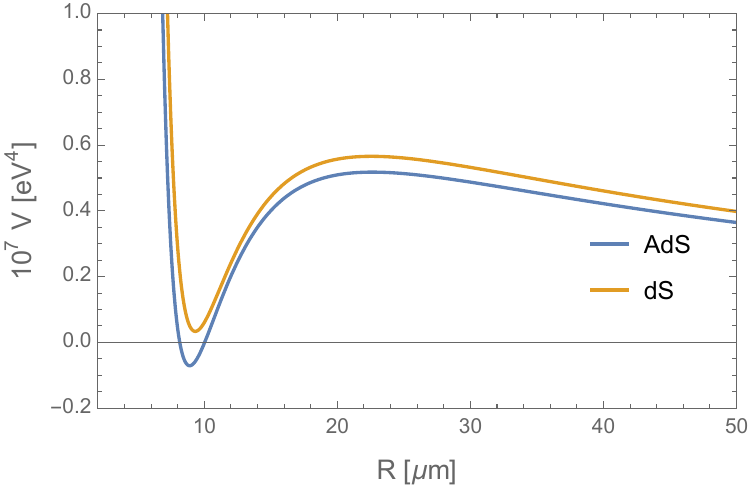}{0.905} 
  \end{minipage}
\begin{minipage}[t]{0.49\textwidth}
\postscript{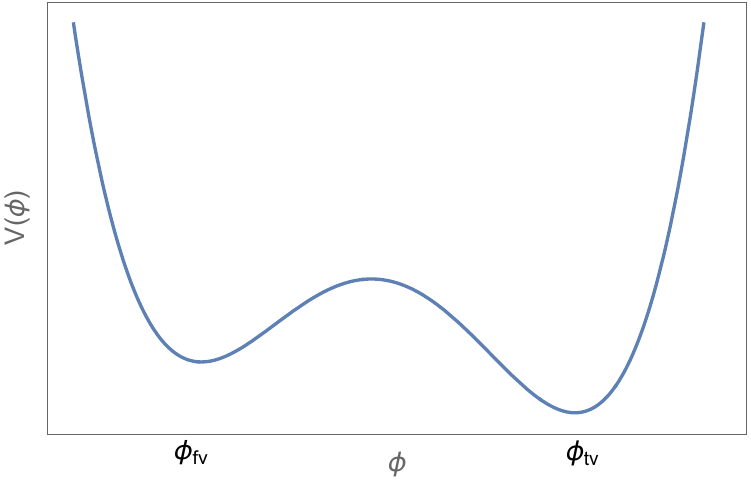}{0.9}
 \end{minipage}
  \caption{{\it Left.} The potential $V(R)$  for $(\Lambda_5)^{1/5} =
    22.6~{\rm meV}$ and $|T_4|^{1/4} =24.2~{\rm meV}$, considering $N_F-N_B = 6$
    (AdS) and $N_F-N_B = 7$ (dS). From Ref.~\cite{Anchordoqui:2023woo}. {\it Right.}  Schematic form of the real scalar potential $V(\phi)$.
   \label{fig:2}}
\end{figure}

Another interesting feature of 5D inflation is that the radion can be stabilized in a local (metastable)
dS vacuum~\cite{Anchordoqui:2023etp}, using the contributions of bulk field
gradients~\cite{Arkani-Hamed:1999lsd} or of the Casimir energy,
assuming a mass for the bulk $R$-handed neutrinos of the same order of
magnitude~\cite{Arkani-Hamed:2007ryu}.
Consider 5D Einstein-de Sitter gravity compactified on a circle $S^1$
endowed with $S^1/\mathbb{Z}_2$ symmetry, and assume that the SM is localized on a D-brane transverse to the compact dimension, whereas gravity spills into the compact space. The 
effective 4D potential of the radion field $R$ is found to be  
\begin{equation}
  V(R) = \frac{2 \pi \ \Lambda_5 \ r^2}{R} +\left(\frac{r}{R} \right)^2 \ T_4  + V_C (R) \,,
\label{V}
\end{equation}
where $\Lambda_5$ is the 5D cosmological constant, $r \equiv \langle
R \rangle$ is the vacuum expectation value of the radion, $T_4$ is the
total 3-brane tension,
and $V_C$ stands for the quantum corrections to the vacuum energy due to Casimir forces. These corrections are expected to become important in the
deep infrared region, because the Casimir contribution to the potential falls off exponentially at
large $R$ compared to the particle wavelength. Indeed, as $R$ decreases different particle thresholds open up, 
\begin{equation}
V_C (R) =  \sum_i \frac{\pi r^2}{32
   \pi^7 R^6} \ (N_F - N_B) \ \Theta (R_i -R) \,,
 \label{xxx}
\end{equation} 
where $m_i =
 R_i^{-1}$ are the masses of the 5D fields, $\Theta$ is a step function, and $N_F - N_B$ stands for the
 difference between the number of light fermionic and bosonic degrees
 of freedom. At the classical level, i.e. considering
 only the first two terms in (\ref{V}), it is straightforward to see that the
 potential develops a maximum at
\begin{equation}
 R_{\rm max} = - T_4/(\pi
 \Lambda_5) \,,
\end{equation}
requiring a negative tension $T_4$. Note that if the fermionic degrees of freedom overwhelm the bosonic contribution, they would give rise to possible
minima, as long as $R_i < R_{\rm max}$. This could be the case if $N_F
= 12$ takes for the three 5D Dirac neutrino fields and $N_B=5$
accounts for the 5D graviton. In Fig.~\ref{fig:2} we show an
illustrative example. 

\section{Living on the Edge: Cosmology within sight of the space boundary}

\subsection{Cosmic Discrepancies}

Over the last few years, low- and high-redshift observations set off
tensions in the measurement of the present-day expansion rate $H_0$
and in the determination of the amplitude of the matter clustering in
the late Universe (parameterized by
$S_8$). More concretely, the values $H_0 =
67.4 \pm 0.5~{\rm km/s/Mpc}$
and $S_8= 0.834 \pm 0.016$ inferred from {\it Planck}'s CMB data assuming $\Lambda$CDM~\cite{Planck:2018vyg} are in
$\sim 5\sigma$ tension with $H_0 =73.04 \pm 1.04~{\rm km/s/Mpc}$ from the SH0ES distance ladder
measurement (using Cepheid-calibrated type-Ia
supernovae)~\cite{Riess:2021jrx,Murakami:2023xuy} and in $\sim 3\sigma$ tension with
$S_8= 0.766^{+0.020}_{-0.014}$ from the cosmic shear data of the
Kilo-Degree Survey (KiDS-1000)~\cite{Heymans:2020gsg},
respectively. These statistically significant discrepancies have
become a new cornerstone of theoretical physics, and many beyond SM
setups are rising to the challenge~~\cite{DiValentino:2021izs,Schoneberg:2021qvd,Perivolaropoulos:2021jda,Abdalla:2022yfr}.

$\Lambda_s$CDM~\cite{Akarsu:2019hmw, Akarsu:2021fol,Akarsu:2022typ,Akarsu:2023mfb} is one of the many beyond SM physics models that have
been proposed to simultaneously resolve the $H_0$ and $S_8$
tensions; see Appendix for details.\footnote{An alternative model that
  accommodates the data has been presented
  in~\cite{Gomez-Valent:2024tdb}.} The model relies on an empirical conjecture which postulates
that $\Lambda$ may have switched sign (from negative to positive) at
critical redshift $z_c \sim 2$;
 \begin{equation} \Lambda\quad\rightarrow\quad\Lambda_{\rm s}\equiv \Lambda_0 \ {\rm sgn}[z_c-z],
\label{Lambdas}
 \end{equation}
 with $\Lambda_0>0$, and where ${\rm sgn}[x]=-1,0,1$ for $x<0$, $x=0$
 and $x>0$, respectively. Apart from resolving the three major cosmological
 tensions, $\Lambda_s$CDM achieves quite a good fit to Lyman-$\alpha$
 data provided $z_c \lesssim 2.3$~\cite{Akarsu:2019hmw}, and it is in agreement with the otherwise puzzling JWST
observations~\cite{Adil:2023ara,Menci:2024rbq}.

Despite the remarkable success of $\Lambda_s$CDM to accommodate the experimental
data, the model is theoretically
unsatisfactory because it postulates that
the Universe experienced a rapid transition
from an AdS vacuum to a dS vacuum, and
this hods out against the AdS-DC conjecture, which posits that flat space limit is at infinite distance in the space of metric
configurations and therefore these two vacua are an infinite distance appart in
metric space~\cite{Lust:2019zwm}.  However, it is important to stress that this no-go theorem is valid at zero temperature, where the
number of light particles is (in general) constant. At finite temperature, particles can decay and hence
the number of light particles can change. In this way the minima of
the potential can be lifted~\cite{Anchordoqui:2023woo}. 

\subsection{AdS $\to$ dS transition driven by Casimir forces of bulk fields}

A possible explanation
for the required AdS $\to$ dS crossover transition in the vacuum
energy can be obtained using the Casimir forces of fields inhabiting
the dark dimension~\cite{Anchordoqui:2023woo}. We assume that the 5D
spectrum contains a light real scalar field 
$\phi$, in addition to the graviton and the three neutrino generations. We further assume that the real scalar has a potential
with two local minima with very small difference in vacuum energy and
bigger curvature (mass) of the
lower one, see Fig.~\ref{fig:2}. At $z_c$ the false vacuum
``tunnels'' to its true vacuum state. After the quantum tunneling
$\phi$ becomes more massive and its contribution to the Casimir energy
becomes exponentially suppressed. The idea here is that for $z \gtrsim
z_c$, we have $N_B=6$, whereas for $z \lesssim z_c$, we have $N_B =
5$. Taking $N_{F} = 12$ to account for the three Dirac neutrino fields
we can use (\ref{V}) in combination with (\ref{xxx}) to obtain an expression for the
effective 4D radion potential. In Fig.~\ref{fig:2} we show an illustrative example of the
AdS $\to$ dS transition produced by $N_F-N_B = 6$ for $z  \gtrsim
z_c$, and $N_F-N_B =5$ for $z \lesssim z_c$. It is important to note that the 5D vacuum transition creates a $\delta V$ contribution to
$\Lambda_5$, where $\delta V = V(\phi_{\rm fv})- V (\phi_{\rm tv})$
corresponding to the vacuum energies of the upper (false vacuum) and the
lower (true vacuum) minima. We have taken $\delta V \ll \Lambda_5$ so
that it does not perturb the analysis producing the curves shown if Fig.~\ref{fig:2}.

Now, the AdS $\to$ dS transition shown in Fig.~\ref{fig:2} slightly deviates from the model
analyzed in~\cite{Akarsu:2023mfb}, because the fields characterizing the deep
infrared region of the dark sector contribute to the effective number
of relativistic neutrino-like
species $N_{\rm eff}$~\cite{Steigman:1977kc}. Using conservation of entropy, fully
thermalized relics with $g_*$ degrees of freedom contribute
\begin{equation}
  \Delta N_{\rm eff} = g_* \left(\frac{43}{ 4g_s}\right)^{4/3} \left
    \{ \begin{array}{ll} 4/7 & {\rm for  \ bosons}\\ 1/2 & {\rm for \
                                                           fermions} \end{array}
               \right.                                        \,,
\end{equation}
where $g_s$ denotes the effective 
degrees of freedom for the entropy of the other thermalized
relativistic species that are present when they decouple~\cite{Anchordoqui:2019amx}. The 5D
graviton has 5 helicities, but the spin-1 helicities do not have zero
modes, because we assume the compactification has
$S^1/\mathbb{Z}_2$ symmetry and so the $\pm 1$ helicities are
projected out. The spin-0 is the
radion and
the spin-2 helicities form the massless (zero mode) graviton. This means
that for the 5D graviton, $g_*=3$. The scalar field $\phi$ contributes with $g_*=1$. The (bulk) left-handed neutrinos are odd, but the right-handed neutrinos are even and so each
counts as a Weyl neutrino, for a total $g_* =2 \times 3$. Assuming that the
dark sector decouples from the SM sector before the electroweak phase
transition we have $g_s = 106.75$. This gives $\Delta N_{\rm eff} =
0.25$.\footnote{It was recently noted that if the QCD axion is
  localized on the SM brane, a combination of theoretical and
  observational constraints forces it to have decay constant in a
  narrow range $10^9 \lesssim f/{\rm GeV} \lesssim 10^{10}$~\cite{Gendler:2024gdo}. This corresponds to a mass
  for the QCD axion of $1 \lesssim m_a/{\rm meV} \lesssim 10$. Although the
  axion would not affect the Casimir corrections to the potential, it
would  contributes to $\Delta N_{\rm eff}$~\cite{Baumann:2016wac,Green:2021hjh,DEramo:2022nvb}.} A numerical study shows that the addition of extra
relativistic degrees of freedom does not spoil the resolution of the
$H_0$ and $S_8$ tensions~\cite{Anchordoqui:2024gfa}.

We end with an observation: the argument to understand the
transition is
essentially the same than the one in finite temperature models, because the number of light degrees of freedom
changes due to a different transition of the 5D scalar field. In plain
English, the model avoids finite temperature requirements and relies on
an ordinary vacuum decay in five dimensions. This obviously implies that the AdS
vacuum is not a true vacuum. The vacuum in the radius modulus is
determined by the contribution to the Casimir potential of the number
of light degrees of freedom. This number changes discontinuously due
to an ordinary vacuum decay of a 5D scalar field which satisfies the AdS-DC conjecture. This
change drives the AdS to dS transition in the radius modulus, which is therefore discontinuous as in first order transitions.

\section{The Black Hole Transition Conjecture}

\begin{figure}[htb!]
  \begin{minipage}[t]{0.49\textwidth}
    \postscript{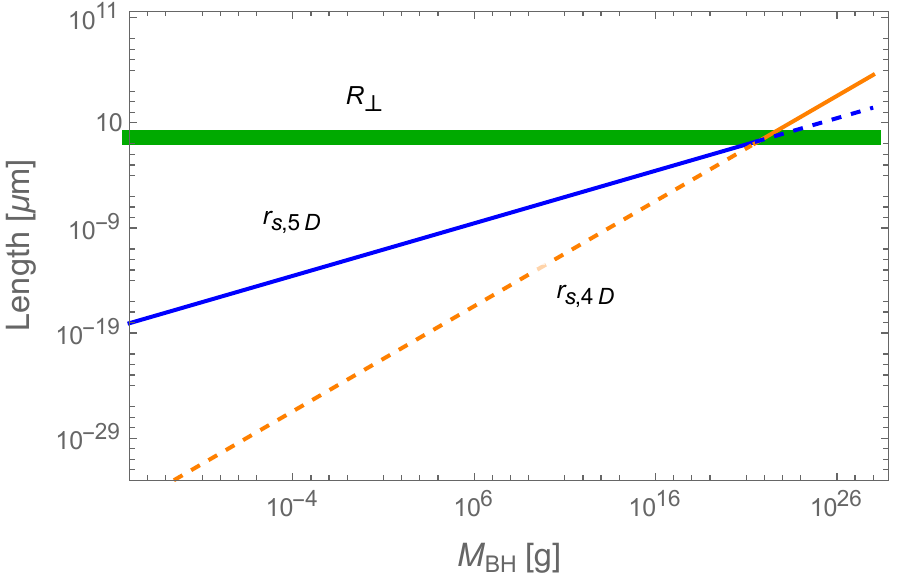}{0.9}
  \end{minipage}
\begin{minipage}[t]{0.483\textwidth}
    \postscript{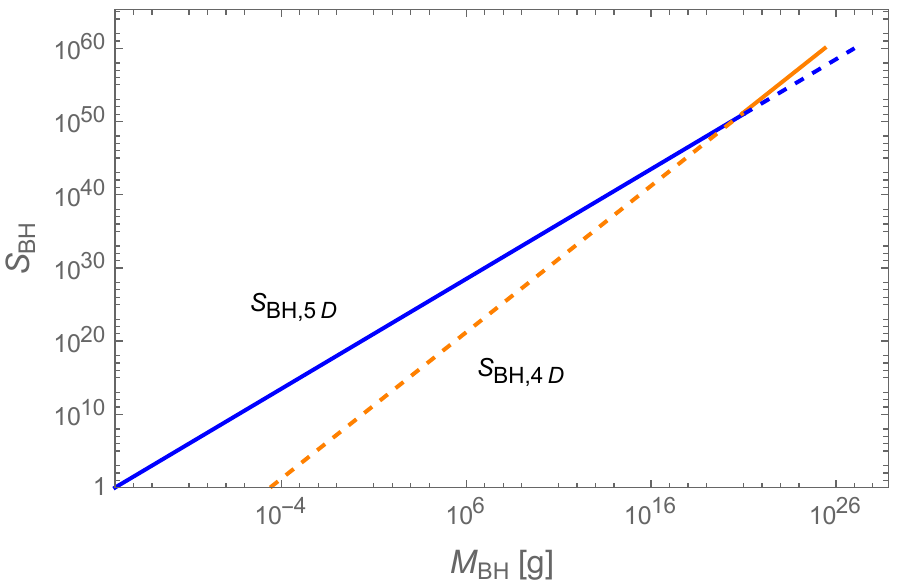}{0.9}
  \end{minipage}
  \caption{Scaling of the Schwarzschild radius (left) and black hole
    entropy (right) in $d=4$ and $d=5$
  dimensions. From Ref.~~\cite{Anchordoqui:2024dxu}. \label{fig:3}}
\end{figure}

In Sec.~\ref{sec:4i}, we have argued that if $r_s < R_\perp$, then 
black holes are 5D, where $r_s$ is the Schwarzschild
 radius and $R_\perp$ the radius of the dark dimension. This implies that if the horizon size of a 4D
black hole that is evaporating shrinks below the micron scale, then
the black hole must undergo a 4D $\to$ 5D transition. As noted
in~\cite{Anchordoqui:2024dxu}, the black hole transition is instantly  
visible by analyzing the black hole entropy of a $d$-dimensional black
hole,
\begin{equation}
S_{\rm BH} = \frac{4 \pi \ M_{\rm BH} \ r_s}{d-2} \sim \left(\frac{M_{\rm
      BH}}{M_d}\right)^{(d-2)/(d-3)} \,,
\label{aa}
\end{equation}  
where $M_{\rm BH}$ is the black hole mass and $M_d$ is the $d$-dimensional Planck scale; note that
for $d=4$ we have $M_4 = {M}_p$. In Fig.~\ref{fig:3} we show a comparison of the 4D and 5D scaling
behavior of the black hole entropy as given by (\ref{aa}). By adding species in a
higher dimensional theory, it follows from (\ref{aa}) 
that  the scaling behavior of
the entropy changes, and for the black hole it is more convenient to be
in the 5D configuration because for given black hole mass its entropy is larger than the one in the 4D
configuration. As can be seen in Fig.~\ref{fig:3}, for $R_\perp \sim 1~\mu{\rm m}$, the transition takes
place at $M_{\rm BH} \sim 10^{21}~{\rm g}$.  Note that the 4D and 5D
entropies as given by (\ref{aa})  are equal at the 5D-4D transition
point where $M_{\rm BH} \sim M_p^2 R_\perp$. Moreover, the entropy crosses the
horizontal axis 
where the black hole masses are the same as the 4D or 5D Planck masses. Then, the associated lengths are the
4D or 5D Planck lengths, where the two entropies are equal to one.

The transition between 4D and 5D black holes corresponds to Gregory-Laflamme phase transition~\cite{Gregory:1993vy} and 
  is also visible using the free energy in terms of the temperature
  $T_H$~\cite{Bedroya:2024uva}. This can be seen in the left plot  in Fig.~\ref{fig:3}, where the black hole mass is approximately the same as the free energy, and the black hole radius corresponds to the inverse temperature.
  So the 5D configuration  for given black hole temperature has  smaller free energy  than   the 4D
configuration.

  More generally, this has lead in~\cite{Bedroya:2024uva} to the
  {\bf black hole transition conjecture} which states that any consistent EFT description of $d$-dimensional QG must exhibit
three scales across its moduli space $\mathscr{M}_{\rm QG}$: {\it
  (i)}~the $d$-dimensional Planck scale $M_d$, which controls the
strength of the Einstein term;
{\it (ii)}~the species scale
$\Lambda_{\rm sp}$, where the higher order gravitational corrections
become important; and {\it (iii)}~the black hole scale $\Lambda_{\rm
  BH}$, where at this temperature, the black hole predicted by EFT
undergoes a phase transition to a more stable solution. Furthermore, $\Lambda_{\rm BH} \lesssim \Lambda_{\rm sp} \lesssim M_d$
everywhere in $\mathscr{M}_{\rm QG}$ and $\Lambda_{\rm BH}$ approaches
the mass scale of the lightest tower at large distances in field
space. So for a KK tower one asymptotically  has that $\Lambda_{\rm BH}=m_{\rm KK}$.
For the line interval with EW9-branes attached
at each end~\cite{Schwarz:2024tet}, the three scales are $m_{\rm KK}
\ll M_{\rm KK}  \lesssim M_{11}$~\cite{Caceres:1996is}.\footnote{Using universality of black hole thermodynamics and properties of gravitational scattering amplitudes it has been shown in~\cite{Bedroya:2024ubj} that some intrinsic features of the density of one-particle states above the minimum black hole mass are an indicator for the existence of large extra dimensions, and cannot be reproduced by any lower-dimensional field theory with finitely many fields satisfying the WEC.}

Let us also add a remark about the definition of the species scale by the minimal black hole in view of the phase transition from the $d$-dimensional, e.g. 4-dimensional, black hole to the higher-dimensional, e.g. 5-dimensional, black hole solution.
Recall that the species length $l_{\rm sp}=1/\Lambda_{\rm sp}$ corresponds to the minimal possible Schwarzschild radius of the BH solution in the EFT.
First, from the higher dimensional perspective, the species length corresponds to the higher-dimensional Planck length,  and this is the scale, where the entropy of the higher dimensional black hole becomes one.
However, 
 also the  lower dimensional perspective makes sense for the definition of the species length,
 i.e. the  species scale formula can be seen entirely from a $d$-dimensional perspective. 
 Namely $l_{\rm sp}$ corresponds to the Schwarzschild radius, where the entropy of the corresponding minimal $d$-dimensional black hole is equal to $N$, the number of species. Using (\ref{SBHmin}) one then obtains $l_{\rm sp}={N^{1\over d-2} M_d^{-1}}$. For the species being the KK modes of an $n$-dimensional compact space  one then obtains
 \begin{equation}
 l_{\rm sp}=M_d^{{2-d\over d+n-2}}m_{\rm KK}^{-{n\over d+n-2}}\, .
 \end{equation}
 Replacing $m_{\rm KK}$ by $\Lambda_{\rm BH}$, one obtains a definition of the species length entirely in terms of $n$ and $\Lambda_{\rm BH}$:
 \begin{equation}
 l_{\rm sp}=M_d^{{2-d\over d+n-2}}\Lambda_{\rm BH}^{-{n\over d+n-2}}\, .
 \end{equation}
 It is not difficult to show that this $d$-dimensional definition of the species scale agrees with the higher dimensional one, using the specific form of the higher dimensional entropy formula and requiring the entropy in higher dimensions to be one.

 Finally, we discuss the correspondence (conjectured
 in~\cite{Anchordoqui:2022tgp}) between the graviton-boundstate
 interpretation of 5D black holes and massive KK gravitons as dark
 matter candidates. We first consider a 4D black hole with entropy
 $S_{\rm BH,4d} = N_{\rm tot}$, with 
 $N_{\rm tot} = N_{g,4} N_{\rm sp}$, where  $N_{g,4}$ is the number
 of 4D gravitons in the black hole boundstate and $N_{\rm sp}$ is the
 number of additional species in the black hole boundstate~\cite{Basile:2024dqq}.  These can
 be the KK gravitons, i.e. $N_{\rm sp} =N_{\rm KK}$.  The radius of
 the 4D black hole is $r_{s,4} = N_{\rm tot}^{1/2} l_p$, where
 $l_p = 1/M_p$. For the minimal black hole, whose radius sets the
 species scale, $N_{g,4}=1$, leading to
 $S_{{\rm BH},4d,{\rm min}}= N_{\rm KK}$. The corresponding minimal
 radius is $r_{s, 4, {\rm min}} = N_{\rm KK}^{1/2} l_p$. This is
 indeed the species length, being just the 5D Planck length.

Now, we consider a 5D black hole after the transition. There are no KK
gravitons anymore, since they are part of normal 5D gravitons.
The black hole is now made entirely of $N_{g,5}$ 5D gravitons and its
entropy becomes $S_{{\rm BH},5d}= N_{g,5}$.
The radius of the 5D black hole is $r_{s,5} = N_{g,5}^{1/3} l_5$,
where $l_5 = 1/M_5$ is the 5D Planck length. For the minimal black hole, whose radius again sets the species scale,
$N_{g,5}=1$, leading to $S_{{\rm BH},5d,{\rm min}}= 1$. The
corresponding minimal radius is $r_{s,5, {\rm min}} =  l_5$. This
is again the species length $l_{\rm sp}$,
so we get in a consistent way that $r_{s,5, {\rm min}} = r_{s,4, {\rm min}}$.

In summary, the 4D/5D phase transition for the KK gravitons can be
understood assuming that above the KK scale these states are
part of the 5D graviton and therefore  the effective description of
space time becomes 5D. So in this case we indeed would have a 5D (or
$d$-dimensional) black hole graviton boundstate description, just like
in the original black hole $N$-portrait picture
of~\cite{Dvali:2011aa}. This implies that before the phase transition,
the correspondence is that the 4D black hole is a bound state of the
$N_{\rm KK}$ particles, whereas after
the phase transition, there is the correspondence between the 5D black hole and the bound state of the 5D gravitons.

\section{Concluding Thoughts}

We have seen that the dark dimension scenario provides one possible
explanation of the
cosmological hierarchy problem and carries with it a rich phenomenology:
\begin{itemize}[noitemsep,topsep=0pt]
\item It provides a profitable arena to accommodate a very
  light gravitino.
\item It encompasses a framework for primordial black
holes,  KK
gravitons, and a fuzzy radion to emerge
 as viable candidates to comprise some or all of the dark matter.
\item It also encompasses an interesting framework for studying cosmology and
astroparticle physics.
\item  It provides a
natural set up for $R$-neutrinos propagating in the
bulk to accommodate neutrino masses in the range
$10^{-4} < m_\nu/{\rm eV} < 10^{-1}$, despite the lack of any
fundamental scale higher than $M_*$. The suppressed neutrino
masses are not the result of a see-saw mechanism, but rather because
the bulk modes have couplings suppressed by the volume of the dark
dimension (akin of the weakness of gravity at long distances).
\end{itemize}
We have also seen that uniform 5D inflation can relate the causal size of the
observable universe to the present weakness of gravitational
interactions by blowing up an extra compact dimension from the
microscopic fundamental length of gravity to a large size in the
micron range, as required by the dark dimension scenario. Moreover,
uniform 5D inflation can lead to an approximate scale invariant power
spectrum of primordial density perturbations. The predicted small-angle ($< 10^\circ$) CMB power spectrum is compatible with observations. Such an angle corresponds to a distance $\sim 2.3~{\rm Mpc}$ and multipole moment $\ell \simeq 30$. For smaller $\ell$ multipoles (larger angles), one obtains more power spectrum than standard 4D inflation, corresponding to a nearly vanishing spectral index, that the present data cannot distinguish due to large errors.  One caveat here is that the power spectrum is cosmic variance limited~\cite{Larson:2010gs}. However, even though cosmic variance prevents identification at low $\ell$, the transition region could provide a signal for experiments in the near future. To determine $\ell$ multipoles in the transition region, an exhaustive transfer-function analysis (numerically solving the linearized Einstein-Boltzmann equations) would be required. The tensor-to-scalar ratio is also
consistent with observations. An estimate of the magnitude of
isocurvature perturbations based on entropy perturbations indicates
that they are suppressed~\cite{Antoniadis:2023sya}. A dedicated investigation along these lines is obviously important to be done.

On a separate track, it was noted in~\cite{Paraskevas:2024ytz} that the cosmic scale factor $a$ describing the evolution of $\Lambda_s$CDM is continuous and non-zero at $t=t_c$, but its first
derivative $\dot a$ is discontinuous, and its second derivative
$\ddot a$ diverges. In the spirit
of~\cite{Akarsu:2024qsi}, it would be interesting to investigate the
evolution of (\ref{V}) during the
phase transition induced by the Casimir forces. This would allow a
complete description of the background and perturbation evolution at
all redshifts.

\section*{Acknowledgements}
We thank Eleonora Di Valentino and Andriana Makridou for permission to
reproduce Figs.~\ref{fig:1} and \ref{fig:4}.  The work of L.A.A. is supported by the U.S. National
Science Foundation (NSF Grant PHY-2112527). I.A. is supported by the Second Century Fund (C2F), Chulalongkorn University. The work of D.L. is supported by the Origins
Excellence Cluster and by the German-Israel-Project (DIP) on Holography and the Swampland.

\section*{Appendix}

\begin{figure}[htb!]
   \begin{minipage}[t]{0.32\textwidth}
    \postscript{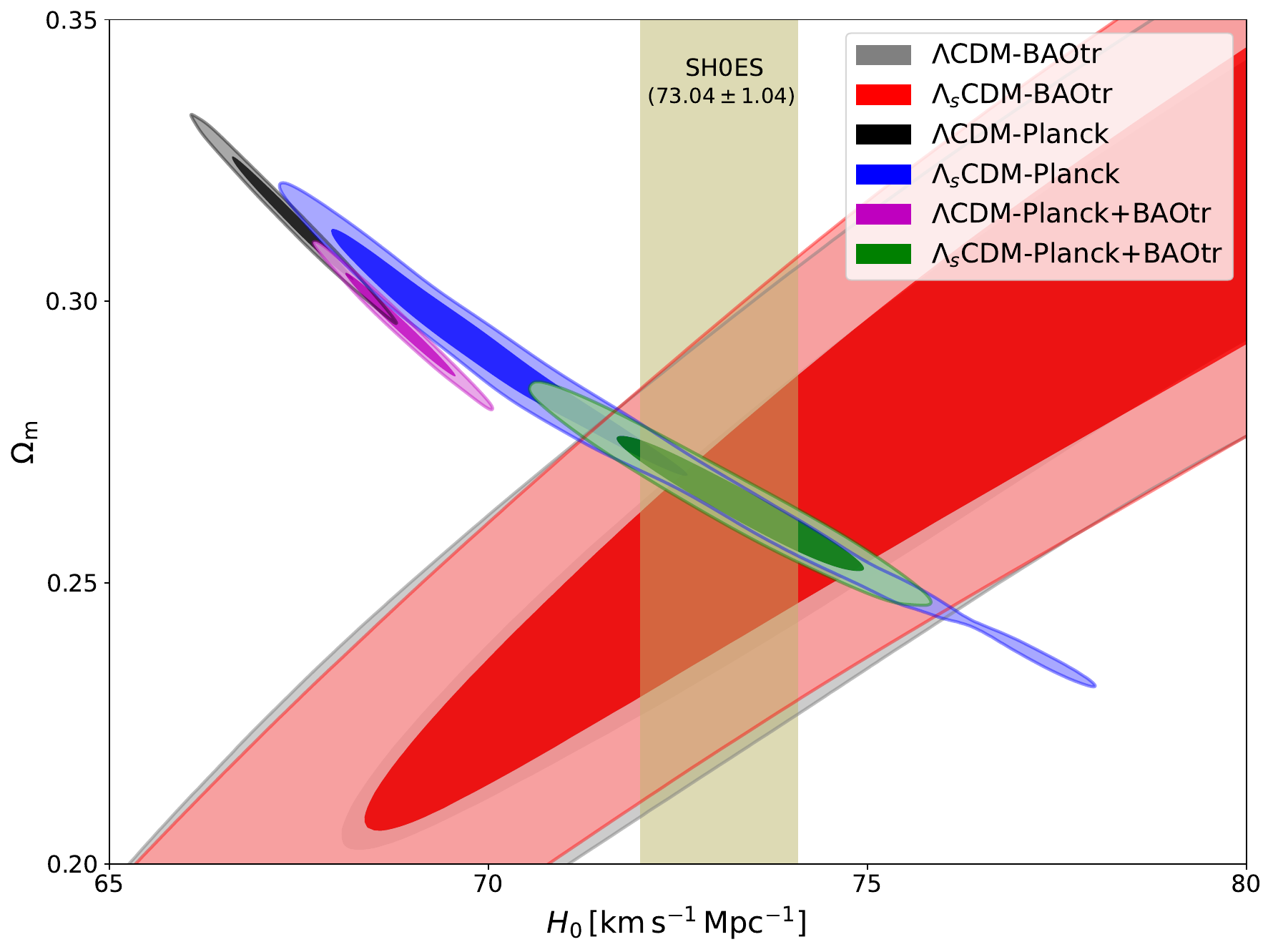}{0.99}
  \end{minipage}
\begin{minipage}[t]{0.33\textwidth}
    \postscript{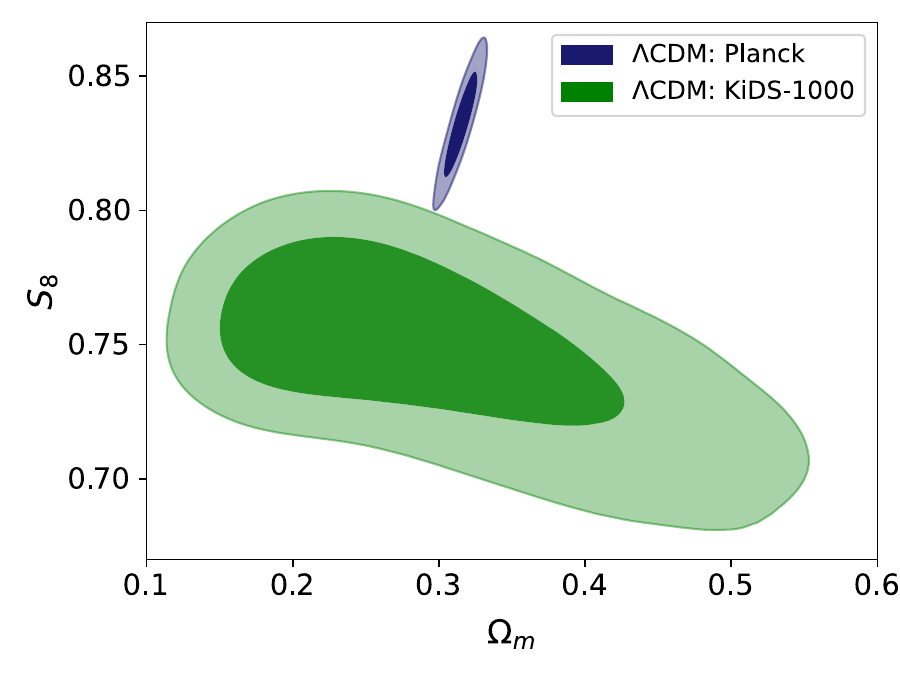}{0.99}
  \end{minipage}
\begin{minipage}[t]{0.33\textwidth}
    \postscript{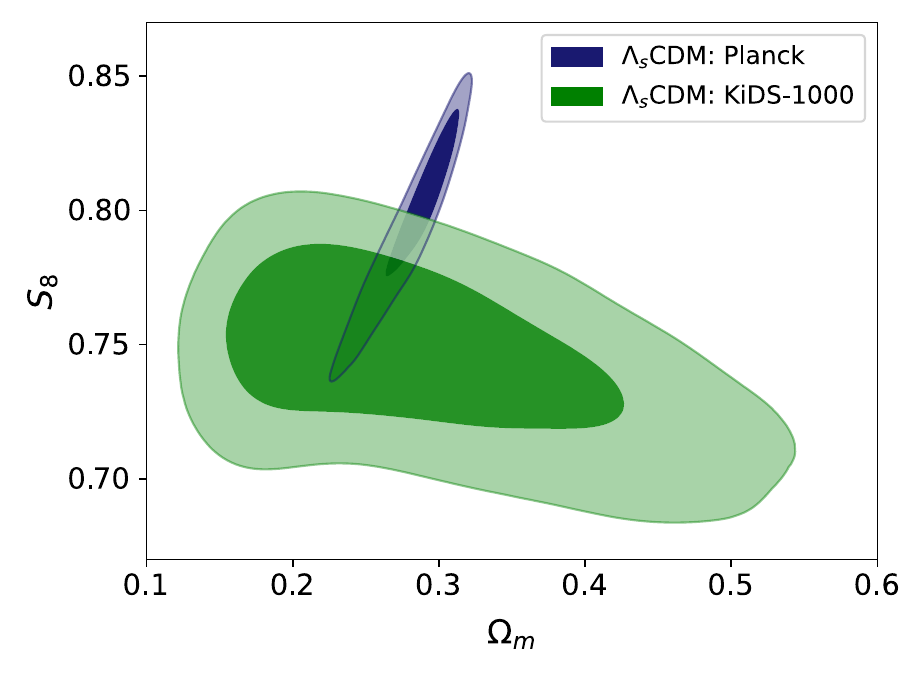}{0.99}
  \end{minipage}
   \caption{2D contours at 68\% and 95\% CL in the $H_0$-$\Omega_m$
     plane (left) for the $\Lambda$CDM and $\Lambda_s$CDM models, and
     in the $\Omega_m$-$S_8$
     plane for the $\Lambda$CDM (middle) and $\Lambda_s$CDM (right)
     models. In the left pannel the model predictions have been
     extracted fitting data from the {\it Planck} satellite and transversal 
baryon acoustic oscillations (BAOtr). In the middle and right panels
the model predictions have its origins in {\it Planck} and KiDS-1000
data. Note that the Planck and BAOtr contours intersect exactly at the
vertical band of the SH0ES measurement. From Ref.~\cite{Akarsu:2023mfb}.}
\label{fig:4}
\end{figure}

In this Appendix we briefly discuss the $\Lambda_s$CDM solution of the
cosmic discrepancies, paying special attention to the role played by
data from
baryon acoustic oscillations (BAO). The most recent analysis
  presented in~\cite{Akarsu:2023mfb} is based on: the {\it Planck} CMB
  data~\cite{Planck:2019nip}, the Pantheon+ supernovae type 
  Ia sample~\cite{Brout:2022vxf}, the data release of
  KiDS-1000~\cite{KiDS:2020suj}, and the (angular) transversal 2D BAO
  data on the shell~\cite{Nunes:2020hzy}, which are
  less model dependent than the 3D BAO data used in previous studies
  of $\Lambda_s$CDM. The outcome, which is displayed in
  Fig.~\ref{fig:4}, shows that the $\Lambda_s$CDM model can
  simultaneously resolve both the
  $H_0$ and $S_8$ tensions. It is important to stress that the BAO
3D data sample assumes $\Lambda$CDM to determine the distance to the
spherical shell, and hence could potentially introduce a bias when
analyzing beyond $\Lambda$CDM models~\cite{Bernui:2023byc}.

As shown in~\cite{Gomez-Valent:2023uof}, 3D BAO data leave no room for low-$z$ solutions to the $H_0$ tension
if the absolute magnitude of supernova $M$ is constant. This is generally phrased as a no-go theorem which
states that if $M$ is constant, to address the $H_0$ tension one needs to consider some sort of new physics at $z>1000$. If one uses 2D BAO (instead of 3D BAO) data, though, it is
possible to solve the Hubble tension without requiring new physics
before recombination. However, in this case the effective dark energy
density needs to be negative at $z\gtrsim 2$ in order to produce the
correct angular diameter distance to the last scattering surface.

\end{document}